\pdfoutput=1

\documentclass[preprint,12pt,authoryear]{elsarticle} 
\pdfoutput=1
\usepackage{siunitx} 
\usepackage{natbib} 
\usepackage{graphicx}
\usepackage{tikz,graphics,color,fullpage,float,epsf,caption,subcaption}

\usepackage{algorithm}
\usepackage{algorithmic}

\usepackage{amssymb}
\usepackage{amsmath}

\usepackage{textcomp}

\newtheorem{remark}{Remark}

\journal{Annual Reviews in Control}

\begin{document}

\begin{frontmatter}




\title{Modeling Cyber-Physical Human Systems via an Interplay Between Reinforcement Learning and Game Theory}


\author[label2]{Mert Albaba}
\author[label2]{Yildiray Yildiz}


\address[label2]{Department of Mechanical Engineering, Bilkent University, Turkey}

\begin{abstract}
Predicting the outcomes of cyber-physical systems with multiple human interactions is a challenging problem. This article reviews a game theoretical approach to address this issue, where reinforcement learning is employed to predict the time-extended interaction dynamics. We explain that the most attractive feature of the method is proposing a computationally feasible approach to simultaneously model multiple humans as decision makers, instead of determining the decision dynamics of the intelligent agent of interest and forcing the others to obey certain kinematic and dynamic constraints imposed by the environment. We present two recent exploitations of the method to model 1) unmanned aircraft integration into the National Airspace System and 2) highway traffic. We conclude the article by providing ongoing and future work about employing, improving and validating the method. We also provide related open problems and research opportunities. 

\end{abstract}


\end{frontmatter}


\section{Introduction}
\label{sec:intro}
In a 2006 NASA report, allocation of tasks, as well as switching, between humans and automation is stated as one of the \textit{highest priority research needs} for a successful next generation airspace development, where several new automation components are expected to be introduced to cope with the inevitable increase in traffic density \citep{Sheridan:06}. A 2012 U.S. Department of Defense report \citep{DoD:12} declares that the taxonomy established by the ``levels of autonomy'' creates ``a focus on machines, rather than on the human-machine system," which in turn ``has led to designs that provide specific functions rather than overall resilient capability". The same report suggests that \textit{human-system collaboration} should be the defining theme for the design and operation of autonomous systems. In a 2017 review article \citep{LAMNABHILAGARRIGUE20171}, where an in-depth analysis is provided about the current and future roles of the systems and control field, the question ``how to optimally conjugate automated systems with the interplay of humans?" is posed as one of the grand challenges. It is therefore clearly understood by engineers and scientists that considering humans as integral parts of complex physical systems that embody communication, computation, control and networking technologies, can potentially accelerate the advancement of technology that can address pressing human needs. This requires a new perspective that considers the human, the physical plant and enabling cyber-technologies as a single system, namely a cyber-physical and human system (CPHS). This is in contrast to imagining the human solely as a user who is isolated from the technology. 

Adopting the CPHS framework brings its own challenges, especially when it comes to obtaining predictive models. Apart from the intricacies of cyber-physical system (CPS) modeling, the human element is usually the most demanding component of a CPHS in terms of forecasting the future behavior. The difficulty intensifies when the system contains more than one human, which requires factoring in multiple human-human and human-automation interactions. The focus of this paper is on the latter type, where human interactions are an inseparable feature of the system. To simplify the exposition in this paper, we use ``human-machine interactions'', ``human-autonomy interactions'' and ``human interactions'' interchangeably, although a more careful use of the language that pays attention to nuances is possible. To obtain realistic models of CPHS, therefore, we need to concentrate on modeling methods that give us the ability to include typical human characteristics. For example, human interactions are generally not deterministic in nature, which can be captured in CPHS models by utilizing a probabilistic modeling framework. Furthermore, before taking an action, a human generally contemplates other intelligent agents' (such as other humans or automation) possible actions and then tries to choose a move that will increase the chances of obtaining the best outcome reflecting his or her preferences. A representational CPHS model needs to incorporate this ``strategic behavior'' \citep{camerer2011behavioral} of humans. Finally, as much as we want to believe otherwise, humans do not always, if at all, act in an optimal fashion. Our cognitive capabilities and computational powers are not always ample enough to provide the best response to a given situation. This final point is important for distinguishing CPHS models from autonomy models, where, based on available information, an algorithm can possibly be designed to react in the most appropriate way. Models that comprise these three attributes of human reactions, namely, being probabilistic, strategic and non-optimal, have a higher chance of success in terms of representing real-life behavior, for the cases where multiple humans are involved.

A solution for CPHS modeling that addresses human interactions is proposed by the introduction of the ``Semi Network-Form Games'' (SNFG) formalism \citep{LeeWolpert:11, lee2013counter}. SNFG merges three modeling tools: Bayesian networks, game theory (GT) and reinforcement learning (RL). While the Bayesian networks form the probabilistic foundation of the method, game theory provides the required mechanism to produce strategic behavior expected from human interactions, and reinforcement learning enables obtaining time-extended scenarios where humans take successive actions. Non-optimal behavior, which is thought to be a typical human trait, emerges naturally in SNFG with the type of the exploited game theoretical solution technique and reinforcement learning. 

The first research result that exploits the SNFG idea of merging GT and RL to create a CPHS modeling framework for a realistic engineering system, including more than 2 humans, has appeared in \cite{MusaviJGCD:16}, where the problem of integrating unmanned aircraft systems (UAS) into the national airspace (NAS) is carefully studied. The results reported in \cite{MusaviJGCD:16} are supported by extensive simulation capabilities, where the interactions of 180 manned aircraft pilots are modeled. This study builds upon earlier initial attempts to model smaller airspace scenarios with human interactions \citep{YildizGNC:12, yildiz2013predicting, Yildiz:14}. Simultaneous modeling of a large number of decision making agents (pilots, in this case) in a complex scenario is difficult due to the computational cost. A common approach in the literature is to model a single decision maker, whose actions are of interest, and force the rest to obey certain kinematic and dynamic constraints, to obtain a reasonable model behavior. However, although this approach presents some insight into the dynamics of the system, it is limiting due to grossly simplifying the real-life human interactions. The study conducted by \cite{MusaviJGCD:16} provided, for the first time in the literature, probabilistic outcomes of UAS integration scenarios, where each of the 180 aircraft pilots are modeled using the game theoretical decision making process, simultaneously. In this research, Bayesian Networks are not utilized and the probabilistic decision making is obtained through employing a stochastic reinforcement learning algorithm proposed in \cite{Jaakkola:94}. Recently, this work is extended for the cases where the aircraft can move both horizontally and vertically in a 3-dimensional airspace \citep{musavi20183d}. This extension required a dramatically larger observation space for the pilots which ruled out the possibility of using exact RL methods. To address this issue, Neural-Fitted Q-iteration \citep{riedmiller2005neural, riedmiller2007learning, gabel2011improved} is integrated into the game theoretical framework, which uses neural networks (NN) for compactly estimating the exact state-action values needed by the RL algorithm. 

Another study, exploiting the same approach, conducted by \cite{li2018game} achieved a similar result in the automotive domain, by creating a modeling framework for road traffic consisting of 50 manned vehicles and an autonomous car. This result was a continuation of leading studies conducted by \cite{Oyler:16} and \cite{Li:16}. Similar to \cite{MusaviJGCD:16} and \cite{musavi20183d}, this contribution is also the first in the automotive literature where a large number of decision making drivers are simultaneously modeled using a game theoretical modeling approach. Recently, an extended version of this work, which covers a larger class of interaction scenarios, with the help of a road traffic simulation on a 5-lane highway, is presented in \cite{Albaba:09}, where validation studies by processing real traffic data, which is provided in \cite{Colyar:07}, is conducted. 

In this paper, we first present the basic components of the modeling approach discussed above and then provide the examples of CPHS framework creation, using GT and RL, for engineering systems that can be employed in predicting the  outcomes of having several humans, automation and physical systems interact with each other in extended periods of time. These ideas exist in the literature in a fragmented manner, and by elucidating them in an aggregated form here, we provide a concise single source. The deliberations in this article on employing game theory and reinforcement learning for building CPHS modeling frameworks should benefit control practitioners whose goal is to obtain models of engineering systems where humans are active players. Finally, we discuss
ongoing and future work about the topic, together with open problems that may provide several different research opportunities for the CPHS community. 

The organization of the paper is as follows: In Section~\ref{sec:exist}, we review the existing work. In Section~\ref{sec:basics}, we explain the basic building blocks of the game theoretical model. In Section~\ref{sec:bring}, we show how these blocks are combined together to form the overall modeling approach. In Sections~\ref{sec:air} and \ref{sec:road}, we present the exploitation of the approach to create models of two different engineering systems containing multiple human interactions. We provide a computational complexity analysis in Section~\ref{sec:cc}. In Section~\ref{sec:future} and \ref{sec:open} we discuss ongoing and future work, and related open problems, respectively. Finally, we provide a summary of the article in Section ~\ref{sec:summary}.

\section{Review of existing work}
\label{sec:exist}

In this section, we review research activities looking for the answer to this question: ``How can we predict the outcomes of engineering scenarios involving a cyber-physical human system with not one but several human elements?''. The literature revolving around this salient question, either partially or completely, covers a wide range of engineering realms, which is hard to exhaustively examine in the limits of this article. Since, up until now, there has been two main studies addressing this question by exploiting the game theoretical modeling approach elaborated in this paper, we will focus on areas that are in the scope of these two research efforts: Unmanned aircraft system integration into the national airspace and road transportation. 

\subsection{Unmanned aircraft systems integration into the national airspace}
\label{sec:unmanned}

Although unmanned aircraft systems (UAS) has been attracting increasingly more attention, we have not yet witnessed the maturation of the civil markets. One of the main reasons for this underutilized potential can be attributed to the lack of regular access to the National Airspace System (NAS) \citep{Dalamagkidis:08}. Due to the well-justified risk-averse nature of the aviation industry, advances in developing rules and procedures for UAS integration into NAS are progressing relatively slowly, which results in UAS flying mainly in restricted airspace. Until it is clearly assured that UAS will not pose a danger to the existing air traffic and thus their integration is proven to be safe, routine access to NAS will not be realized \citep{US_UAV:13, European_UAV:13}. There are many studies that address the problem of UAS integration into the airspace in terms of providing methods and tools to ensure safety. \cite{ding2016initial} proposes an autonomous decision making system for UAVs for determining a safe landing site in the case of an anomaly. For UAV platoons, controller design frameworks are presented in \cite{chen2015safe, chen2016multi, chen2017reachability}, where collisions are eliminated and target states are reached. In \cite{chen2017provably}, collision-free trajectories are designed for large-scale multi-UAV systems in the presence of disturbances in vehicle dynamics, and the method is demonstrated using up to 200 UAVs, in simulation environment. 

 UAS integration still remains to be a challenge \citep{melnyk2019demonstration} and since we don't have the necessary experience to evaluate the effects of integration and there is not enough data yet, the only way to predict the outcomes of adding UAS into the airspace is conducting careful simulations \citep{MITRE:14}. 
To obtain reliable simulation results, a high-fidelity model of the airspace in the presence of manned and unmanned aircraft together with their interactions need to be obtained. 

A typical approach in the airspace traffic modeling literature includes the assumption of pilots always following an ideal behavior pattern without any deviations. This is an unrealistic assumption since, as discussed earlier, a representative model should allow probabilistic human behavior. For example, it is well-documented that pilots may ignore controller's commands or do not obey traffic collision avoidance system (TCAS) resolution advisories during emergency situations \citep{pritchett2010system}. In addition, it is reported that only 13\% of pilot reactions agreed with the pilot model, which predicts a deterministic behavior, used for establishing TCAS algorithms \citep{kuchar2007traffic, LeeWolpert:11}.

One of the main issues that needs to be solved for a safe integration is the development of a dependable sense-and-avoid (SAA) technology for the unmanned aircraft systems. It is not possible to mature this technology without testing its performance through simulations with reliable pilot models \citep{Maki:12}. There exist several SAA methods introduced in the literature, with validations conducted via simulations and experiments. \cite{Kuchar:04} carefully analyzed the utilization of TCAS as the SAA logic, and performed simulation tests using the aircraft encounter model developed by \cite{kochenderfer2008correlated}. During the tests, pilot reactions were assumed to be known beforehand, based on the respective motions of the conflicting aircraft. In another study, \cite{Batlle:12} suggested maneuvers to solve UAS separation conflicts and tested these suggestions with simulations that contain pilots following the recommended maneuvers without any error. The performance of the SAA algorithm proposed in \cite{Florent:10} was assessed by simulations and experiments. For these assessments, the manned aircraft in separation conflict with the UAS were assumed to continue their motion unaffected, while the UAS were implementing the proposed SAA technique. Another example where predefined pilot action models are employed for evaluating various SAA algorithms can be found in \cite{Billingsley:06}.

The modeling framework for UAS integration into NAS that is discussed in this article differentiates itself from the above mentioned environments by providing a platform where pilot actions are not pre-determined but obtained through a decision making process by satisfying a utility function, or a ``happiness function'', that reflects pilot preferences. Furthermore, with the proposed approach, several pilot-pilot and pilot-UAS interactions (180 of them) can be modeled in time-extended scenarios, in a probabilistic manner, with the help of the convergence of game theory and reinforcement learning. The details of this framework are provided in Section \ref{sec:air}.

\subsection{Road transportation}
\label{sec:traffic}

We have reliable physical models of road vehicles that have high predictive power. However, the modeling problem quickly becomes difficult to handle when the task is modeling the vehicle together with the driver operating it. Furthermore, if the desired outcome is a model for traffic containing several vehicles, both manned and unmanned, obtaining answers turns out to be a real challenge. The main obstacle in this matter in question is the lack of accurate human interaction models. 

Valid human interaction models in road traffic may prove themselves useful for two main tasks: Creating accurate traffic simulators that can be used for initial testing and tuning of autonomous car control algorithms, and designing autonomous vehicle control systems based on human way of driving, which can improve the passengers' comfort by making them feel as if a human driver is in control \citep{Carvalho:15}. 

There are several successful driver modeling studies in the literature. Real traffic data is employed to obtain Hidden Markov Model based driver models in \cite{ Lefevre:14} and \cite{Lefevre:15}. A semiautonomous vehicle control architecture is proposed in \cite{Vasudevan:12} and \cite{Shia:14}, where the driver model is obtained, through k-means clustering, and used to inform the controller that produces corrections for driver inputs. Logical, if-then-else commands  form the driver decisions in a modeling framework created by \cite{Salvucci:01}. A multi-agent simulator is employed to 
model lane changing in \cite{Hidas:02}. Lane changing behavior is modeled also in \cite{Kumar:13}, where Bayesian filters and support vector machines are utilized to predict driver intent. There is another body of work on modeling drivers as controllers in a closed loop control system, examples of which can be found in \cite{Hess:90, Sharp:00, Treiber:00, Salvucci:04} and \cite{Ungoren:05}. A more recent example of feedback controller type driver modeling can be seen in \cite{wakitani2018design}. 

The proposed modeling framework for road traffic in this article has the following distinctions compared to earlier work: 1) The driver interaction models are scalable and a traffic scenario consisting of several vehicles interacting with each other can be modeled using a game theoretical approach; 2) driver behavior is obtained through employing a decision making process, instead of assuming a preset driver action model that is a function of time or states; 3) driver models are simultaneously strategic, meaning that they consider other intelligent agents (drivers, unmanned vehicles) possible actions and produce a response accordingly, based on a utility function representing their priorities. This is in contrast to modeling a single driver as a decision maker while assigning pre-determined trajectory profiles for the others. All the listed advantages are realized through the game theoretical modeling approach discussed in this article. There are other game theoretical driving modeling approaches reported in the literature such as the ones proposed by \cite{Yoo:12} and \cite{Yoo:13}, where driver interactions are successfully modeled. On the other hand, unlike the method discussed in this article, these studies do not consider a time-extended scenario. \cite{Ilya:14} also consider a game theoretical approach to model the interactions between the driver and the ego vehicle's power train. By penalizing the vehicle-driver system features of fuel consumption, emissions, battery state and operating conditions, this approach is demonstrated, via experiments, to perform better than the baseline controller, in terms of these system features, while still providing good drivability. However, the utilized game theoretical approach, namely Stackelberg solution, quickly becomes computationally intractable as the number of intelligent agents in the game increases, unlike the hierarchical game theoretical approach used in the method proposed in this article. The details of the proposed method are provided in Section \ref{sec:road}.

\section{The basics}
\label{sec:basics}

In this section, we present the fundamental building blocks of the game theoretical modeling framework discussed in this article. These blocks are game theory and reinforcement learning. Below, we explain these pieces with a relatively narrow scope, using a semi-formal language to allow easy access, and with enough details that will enable understanding of the basic ideas necessary to grasp the following sections. There are several sources in the literature that can be used for formal introductions to these topics, such as \cite{fudenberg1991game} and \cite{camerer2011behavioral} for game theory, and \cite{Wiering:12} and \cite{sutton2018reinforcement} for reinforcement learning. 

\subsection{Game theory}
\label{sec:game}

Game theory studies the interactions between strategic agents. A strategic agent is one that considers other agents' possible actions and their effects on the game while making his or her own decisions. The theory makes predictions about the outcomes of these interactions using precise mathematics. 

\begin{figure}[bht]
\centering
 \includegraphics[trim = 66mm 97mm 79mm 40mm, clip, width=0.5\textwidth]{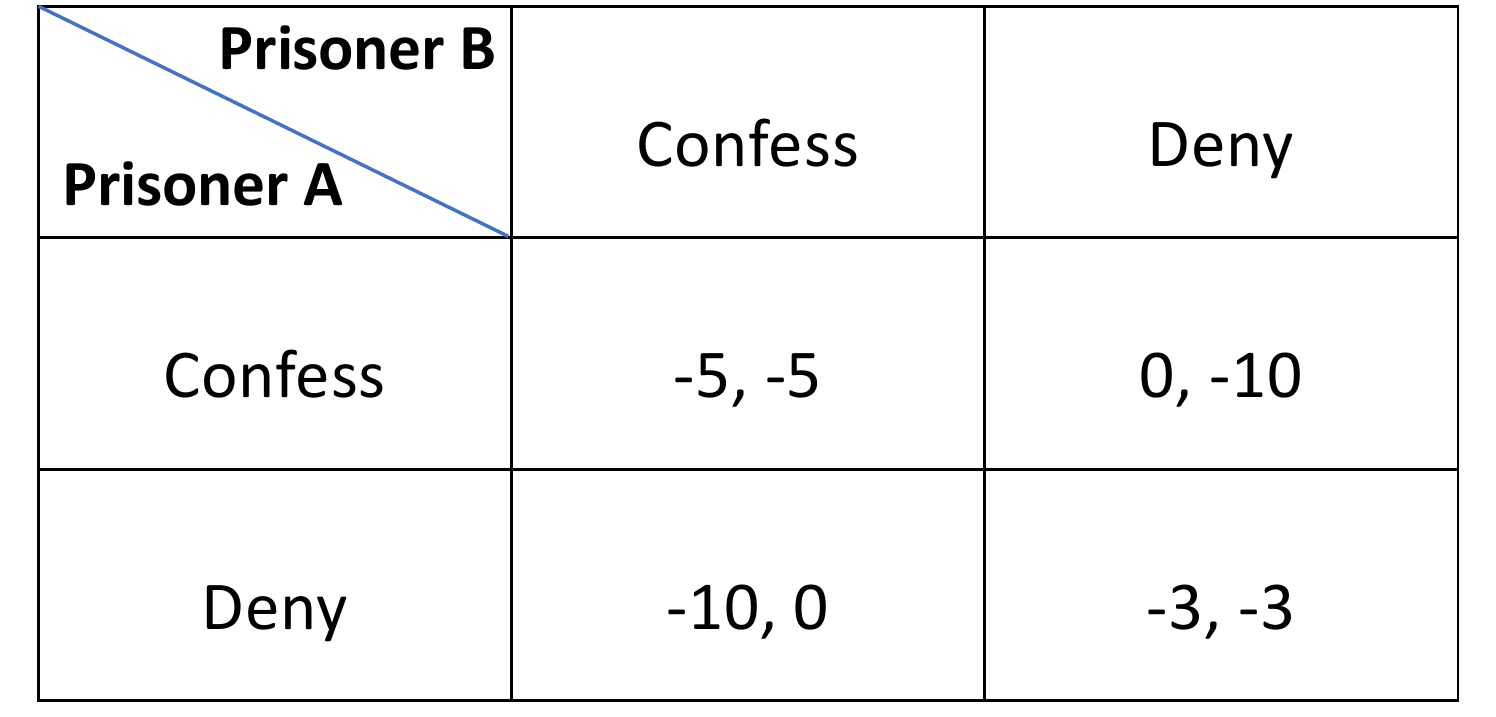}
 \caption{Prisoner's Dilemma}
 \label{f:Prisoner}
\end{figure}

\textit{Players} in a game theoretical setting refer to the entities who can effect the game by their \textit{moves} (or actions, or decisions). \textit{Strategy} of a player defines the procedure based on which a player chooses his or her actions. A \textit{solution concept} is a well established set of rules that are used to predict how a game will unfold. A \textit{Nash equilibrium} is a solution concept, defined similarly to the equilibrium in system dynamics: When players have no incentive to deviate from their selected actions, the game is said to be in Nash equilibrium. This means that in Nash equilibrium, players choose their best actions against each others' actions. A typical example where Nash equilibrium can be observed is a game called the \textit{Prisoner's Dilemma}. In this game, there are two prisoners, Prisoner A and Prisoner B. They are put in separate rooms so that they can not communicate. Both of them are provided with the following information: If Prisoner A confesses the crime, he will be released provided that Prisoner B denies the crime, which will cause Prisoner B serve 10 years in prison. Similarly, if Prisoner B confesses, he will be released provided that Prisoner A denies the crime, which will put Prisoner A in prison for 10 years. If they both deny, each will serve 3 years. If they both confess, each will serve 5 years in prison. We can represent this game in matrix form as shown in Figure~\ref{f:Prisoner}, where players' \textit{payoffs} amount to the negative of the years to be served in prison. It is seen that, although resulting in a low overall payoff, (Confess, Confess) choice is the only Nash equilibrium since no player wants to change their move once they are in this state. It is important to note that Nash equilibrium is not always unique and there may be more than one Nash equilibrium depending on the game. 

There are other equilibrium concepts such as \textit{quantal response equilibrium} where instead of giving the best response to other players' actions, the players choose a probability distribution over their action space where actions with higher expected payoffs have higher probability of being played.

Not all solution concepts predict an equilibrium. For example, \textit{level-k thinking} is a non-equilibrium game theoretical model of strategic interactions, which assigns different levels of \textit{reasoning} for players \citep{Stahl:95, Costa-Gomes:09}. In this model, the lowest level of reasoning is level-0, which represents non-strategic thinking, simply meaning that the players who reason at this level have a strategy that does not take into account other players' possible actions. A level-1 player, on the other hand, takes the best action assuming that his or her opponents are level-0 players. Similarly, a level-k player responds best to his \textit{belief} that the other players are reasoning at level-(k-1). Therefore, this model assumes an \textit{iterated} best response \citep{crawford2008modeling}. Experimental results presented in \cite{camerer2011behavioral} corroborate the predictions of this solution concept with varying success. 

An elementary example for the level-k reasoning model can be given considering two people, named Diana and Ritchie, walking towards each other, along a collision path, in a university corridor. If Ritchie decides to continue walking without considering Diana's possible actions, Ritchie can be considered as a level-0, non-strategic thinker. On the other hand, if Diana believes that Ritchie is a level-0 thinker and therefore decides that the best action is stepping right, then Diana can be modeled as a level-1 player. Although there exists experimental evidence for level-k predictions, this simple example presents a difficulty in this approach: The players' beliefs of others may be wrong. Another issue is that even if the players correctly estimate their opponents' levels, the behavior patterns coded by different levels are sensitive to the selection of the level-0 algorithm. That's why level-0 is often referred as the \textit{anchoring level}.  Although these ``problems'' are real, they may actually be considered as the strengths of this method when it comes to modeling humans. As discussed in the Introduction section, in CPHS with multiple humans, to obtain reasonable predictions, we need models that do not foresee optimal behavior all the time. Another perspective is that leve\mbox{l-k} thinking represents the interactions between intelligent agents that do not have a long interaction history, therefore it is actually expected that their initial assumptions about each others' strategies may not fully reflect reality. This also helps obtain non-optimal human reactions that are providing best responses to their beliefs about the outside world. 

\subsection{Reinforcement learning}
\label{sec:RL}

Reinforcement learning (RL) can be defined as a mathematical representation of learning through reward and punishment. To clarify this definition, and to be able explain the RL algorithm used in the CPHS modeling framework discussed in this paper, we first need to identify main elements of RL and make certain definitions that hold true for almost all RL methods. In RL, there exists an \textit{agent} capable of exerting \textit{actions} that can change the \textit{state} of the environment where the agent operates. The RL problem can be defined as finding the optimal set of action sequence for an agent to achieve a given goal defined as a function of the environment states, through interaction with the said environment. For example, the problem may be making a mobile robot (agent) to go from point A to point B (goal) in a 10 by 10 grid-world with obstacles (environment) by deciding whether to move left, right, forward or backward (actions), in every step of the way. In this scenario, the state can be defined as the grid location the robot is occupying. 

\begin{figure}[bth]
\centering
 \includegraphics[trim = 30mm 80mm 45mm 25mm, clip, width=0.7\textwidth]{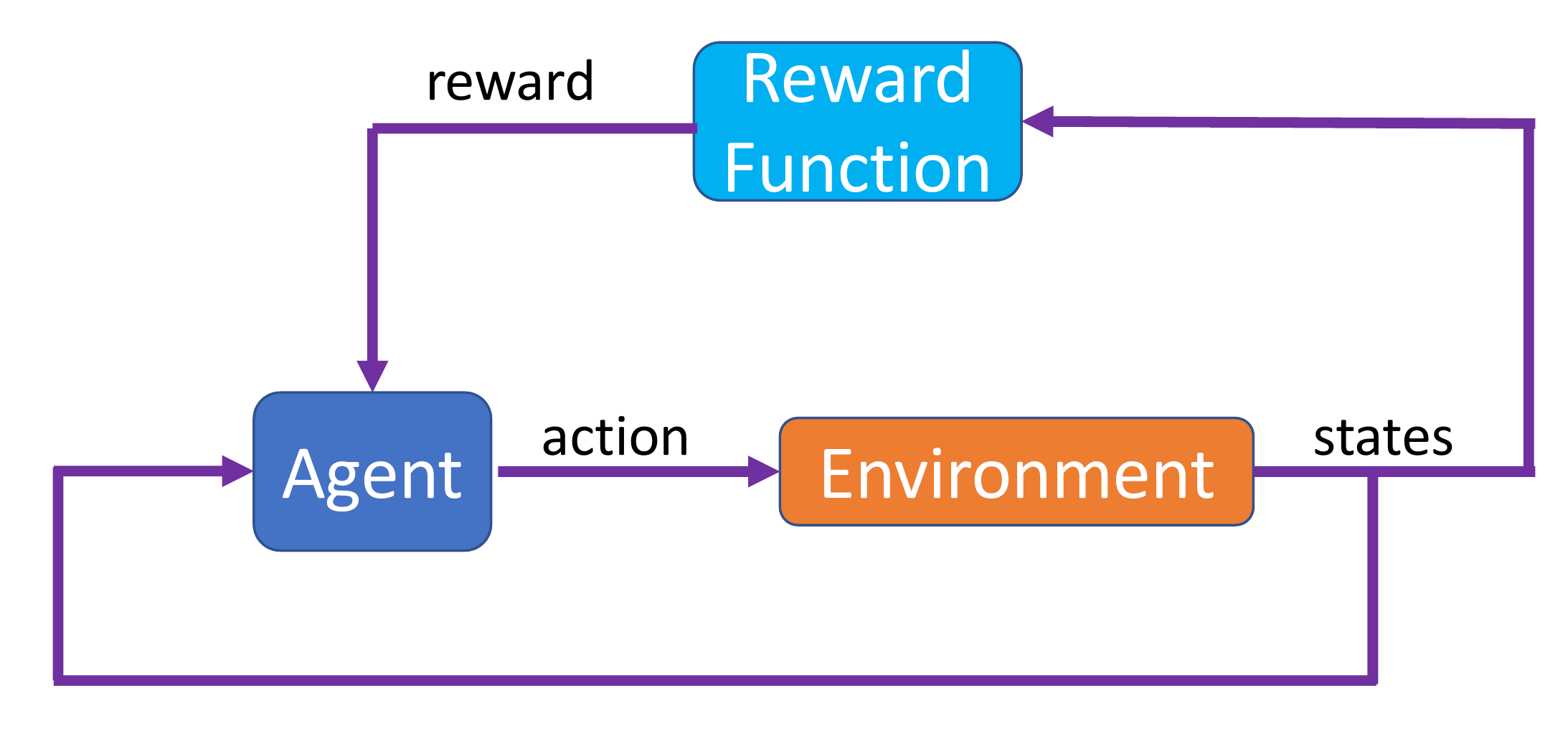}
 \caption{Reinforcement learning process}
 \label{f:RL}
\end{figure}

RL uses the idea of a \textit{reward function} to describe the preferences of the agent (or the designer of the agent) while its learning to achieve a predetermined goal. In the mobile robot example given above, the reward can simply be defined as zero if the robot is at the goal state of point B, and a fixed negative number, otherwise. A \textit{policy} is defined as a probabilistic map from states to actions. The task of the RL algorithm is to find a policy that will make the agent maximize a cumulative discounted reward, during the time of its operation. One way to express the cumulative reward is given as
\begin{equation} \label{cumulative}
C =\sum_{t=0}^{\infty}{\gamma^t r_{t}},
\end{equation}
where $\gamma$ is the discount factor and $r$ is the reward obtained in every step $t$. There are various RL techniques proposed to discover action sequences that will attain the goal of maximizing (\ref{cumulative}). Almost all of these different methods are based on estimating \textit{value functions}, which can be regarded as the value of being in a certain state, based on the policy being implemented. A value function can be given as
\begin{equation}
V^{\pi}(s)=E_{\pi} \Bigg\{ \sum_{k=0}^{\infty} \gamma^kr_{t+k} | s_t=s \Bigg\}
\end{equation}
where $\pi$ represents the implemented policy and $s$ represents state. A similar function, the estimation of which also characterizes the RL method, stands for the value of taking a certain action $a$, in a given state $s$. This function is defined as
\begin{equation}
Q^{\pi}(s,a)=E_{\pi} \Bigg\{ \sum_{k=0}^{\infty} \gamma^kr_{t+k} | s_t=s, a_t=a \Bigg\}.
\end{equation}
The aim of RL is to find the optimum policy $\pi^*$ that will maximize (optimize) the value function. The optimal value function is written as $V^{\pi^*}$ and all its state values are larger than or equal to that of all other value functions that are created by policies different than the optimal policy. This can be represented as $V^{\pi^*}(s) \geq V^{\pi}(s), \forall \pi, \forall s$. Similarly, we can write $Q^{\pi^*}(s,a) \geq Q^{\pi}(s,a), \forall \pi, \forall (s,a)$, for the optimal action value function. Once the optimal action value function is found, the policy 
\begin{equation}
\pi^{*}(s)=\arg \max_a{Q^{\pi^*}(s,a)}
\end{equation}
can be used to select the best action in each state. The answer to the question ``how to find the optimal value function'' determines the type of RL algorithm. The process of finding the optimal policy is sometimes called \textit{training}. Figure~\ref{f:RL} depicts the general training process of RL. In the figure, the agent observes the states and produces an action based on the observed states. This action influences the environment and results in a new set of states. These new states are evaluated by the reward function and a reward signal is formed. The agent uses this signal to update the policy that is being trained and the next cycle starts with the new action. 

In the following subsections, we explain one of the most basic RL algorithms, \textit{Q-learning}, and then present two other RL methods that are utilized in the game theoretic modeling framework elaborated in this article. 

\begin{remark}
The RL algorithms used in this paper are used to obtain human response policies, which can then be used to quantitatively
analyze scenarios where humans are involved, in the simulation environment. Therefore, there is no physical interaction with the environment during learning. 
\end{remark}

\subsubsection{Q-learning}
One of the RL methods that played a significant role for the success of RL is \textit{Q-learning} \citep{watkins1989learning}. In Q-learning, an incremental estimate of the optimal action value function is realized using the update rule
\begin{equation}
Q_{k+1}(s_t,a_t)=Q_{k}(s_t,a_t) + \alpha\bigg(r_t + \gamma \max_a Q_k(s_{t+1},a)-Q_k(s_t,a_t) \bigg),
\end{equation}
where $\alpha$ is the step size and $\gamma$ is the forgetting factor. It is noted that Q-learning is indifferent to the policy that the agent uses during training to \textit{explore} the environment, which means moving from one state to the other . This type of RL algorithms, where exploration and value function updates (or policy updates) are independent, are called \textit{off policy} methods. It can be shown that the learned action value function $Q$ converges to the optimal action-value function $Q^{*}$ with probability 1, if all state-action pairs $(s,a)$ continue to be visited during training, and if the number of these visits converges to infinity. Convergence is achieved exactly if the step size parameter obeys a variant of the stochastic approximation conditions, given as $\Sigma_{k=1}^\infty \alpha_k=\infty$, $\Sigma_{k=1}^\infty \alpha_k^2 \leq \infty$. For a constant step size parameter, convergence is achieved in the mean if $0 \leq \alpha \leq 1$. It is noted that continuously visiting state-action pairs is a requirement for any algorithm that is developed for the general case, where a model for the environment is not provided, to achieve convergence to the optimal solution.\citep{Sutton:98}. The discount factor $\gamma$ determines how valuable the future rewards are. For instance, when $\gamma$ is 0, agent only tries to maximize the immediate rewards. As this value approaches to 1, the agent becomes more far-sighted.

\subsubsection{Neural fitted Q-learning}
\label{sec:nfq}
Instead of keeping a table of Q values, some RL methods use compact structures that provide approximate Q values. This approach is especially useful when the state space is very large. Neural Networks (NN) are one of the effective tools used to store the Q-values in a compact manner, thanks to their universal approximation property. Unlike the conventional Q-learning method explained above, the state-action values are not stored in a table but instead, computed as the output of a function constructed by the specific NN structure: If a state-action pair is fed to the NN as the input, the corresponding approximate Q-value can be obtained as the NN output.  The training of the NN can be achieved by first defining an error function reflecting the difference between the current and the target Q-values, and then minimizing this function by back-propagation. Although calculating the Q-values using a NN provides an effective method, it can fail, either completely or by requiring impractical convergence times, due to the global representation mechanism \citep{riedmiller2005neural}: During the training process, NN weights are updated after the introduction of each individual state-action pair, which also effects the Q-values of other pairs. This may nullify the previous training gains in other regions. On the other hand, the global representation enables the generalization power of NNs by assigning similar Q-values to similar state-action pairs, and thus eliminating the need to train the NN for every possible pair. Therefore, a method is needed that can both exploit this property and eliminate its detrimental effects.

Neural Fitted Q-learning, proposed by \cite{riedmiller2005neural}, achieves to both employ the generalization power of NNs and prevent its potentially harmful effects by storing previous experiences in the form of 3-tuples, $(s,a,s')$, in which $s$ is the original state, $a$ is the action taken and $s'$ is the reached state, and reusing these experiences whenever an update is performed after the introduction of a new data point. Calling a collection of these experiences as set $E$, the NFQ method is given in Algorithm \ref{alg:NFQ}. In the implementation of NFQ learning, it is advised that instead of a random collection of experiences, greedy search, using available Q-values, and random exploration are used together.
{\color{blue}
\begin{algorithm}
	\caption{NFQ Algorithm}
	\begin{algorithmic}[1]
		\STATE  {$k = 0$}
		\STATE  {Initialize the neural network}
		\WHILE  {$k  <  N$}
		\STATE {Generate experience set $G = \{(input^i, target^i), i = 1, 2, ... , \#E\}$ where}
		\STATE {\hspace{\algorithmicindent} $input^i = s^i, a^i$, where $s^i$ is the state and $a^i$ is the action of $i^{th}$ experience,}
		\STATE {\hspace{\algorithmicindent} $target^i = r^i + \gamma \min_{a}Q_k({s'}^i, a)$, where $r^i$ is the transition cost and $\gamma \min_{a}Q_k({s'}^i, a)$ is the weighted expected maximum path reward for the next state ${s'}^i$. }
		\STATE {Calculate the batch error as $\Sigma_{i=1}^n (Q^k(s_i,a_i) - target^i)^2$, where $n$ refers to the experience set size. }
		\STATE {Train the network to minimize the batch error, using resilient back-propagation and obtain $Q_{k+1}$ }.
		\STATE  {$k += 1$}
		\ENDWHILE
	\end{algorithmic} \label{alg:NFQ}
\end{algorithm}}

NFQ contains two types of hyper-parameters: parameter of the Q-learning, the forgetting factor $\gamma$, and the parameters used for NN training. The selection of the Q-learning hyper-parameter $\gamma$ is explained in the previous section. For NN training, the resilient propagation algorithm proposed in \cite{riedmiller1993direct}, which works well for batch learning, is suggested since NFQ is also based on batch learning. It is discussed in   \cite{riedmiller1993direct} that different values of hyperparameters do not effect the performance of the algorithm dramatically and therefore some predefined values suggested in the paper can be used for most of the problems. In many NN training software packages \citep{tensorflow2015-whitepaper, paszke2017automatic} these values are already set as default so no further tuning is necessary.

\subsubsection{Jaakkola reinforcement learning}
\label{sec:Jaak}

An agent being trained by RL uses the available information from the environment. This information is generally called the ``\textit{state}'' of the environment. When the state of the environment contains all relevant information about the current and the past interaction dynamics between the agent and the environment, this state is said to have the Markov property \citep{sutton2018reinforcement}. A learning task involving interactions with an environment that has Markov property is called a Markov Decision Process (MDP). More specifically, defining the probability of transitioning from state ``$s$'' to state ``$s'$'' and obtaining a reward ``$r$'', given an action ``$a$'' as $P(s',r|s,a)$, if this probability depends only on ``$s$'' and ``$a$'' but not on earlier states and actions, this learning task is called an MDP. In almost all RL methods that have convergence guarantees, the underlying dynamics is assumed to be an MDP. In the aerospace and automotive application scenarios that are investigated in this paper, although the underlying dynamics are MDPs, the agents can realistically observe only a portion of the available states. Therefore, from the agents' point of view, the tasks are Partially Observable Markov Decision Processes (POMDP). 

Jaakkola reinforcement learning algorithm \citep{Jaakkola:94} is developed specifically for systems that can be modeled as POMDPs and therefore is a suitable RL method to be employed in the learning tasks that are discussed in this work. In Jaakkola algorithm, along with $Q$-function, the value function, $V$, is also used. At the beginning of the Jaakkola Algorithm, Q values are set to zero for each state-action pair. Moreover, for each state, probability distribution of actions is set to a uniform distribution. Then, for each iteration $(s,a,s')$, $Q$ and $V$ values are updated according to following equations:
\begin{eqnarray}
\beta_t(s,a) &=& \left(1-\frac{\chi_t(s,a)}{K_t(s,a)}\right) \gamma_t\beta_{t-1}(s,a) + \frac{\chi_t(s,a)}{K_t(s,a)} \nonumber \\
\beta_t(s) &=& \left(1-\frac{\chi_t(s)}{K_t(a)}\right) \gamma_t\beta_{t-1}(s) + \frac{\chi_t(s)}{K_t(s)} \nonumber \\
Q_t(s,a) &=& \left(1-\frac{\chi_t(s,a)}{K_t(s,a)}\right)Q_{t-1}(s,a) + \beta_t(s,a)(R_t - R) \nonumber \\
V_t(s) &=& \left(1-\frac{\chi_t(s)}{K_t(s)}\right)V_{t-1}(s) + \beta_t(s)(R_t - R) \nonumber \\
\end{eqnarray}
where, $s$ is the state, $a$ is the action and $t$ is the time step. Moreover, $\chi_t(s,a)$ ($\chi_t(s)$) is equal to 1 if the given state-action pair (state) is visited, and 0 otherwise; $K_t(m,a)$ ($K_t(s)$) is the number of times the state-action pair (state) is visited; $R_t$ is the reward in time step $t$; $R$ is average reward and $\gamma_t$ is the discount factor. After the calculation of $Q$ and $V$ functions, Jaakkola algorithm updates its trained policy $\pi(a|s)$ using the update rule
\begin{equation}
\pi(a|s)=(1-\epsilon)\pi(a|s)+\epsilon\pi^1(a|s),
\end{equation}
where $\epsilon$ is the update rate, and $\pi^1(a|s)$, the policy that the trained policy is being changed towards, is a greedy-policy based on the calculated Q(s,a) values. In other words, $\pi^1(a|s)$=1 if the action ``\textit{a}'' has the highest Q-value in a given state ``\textit{s}''. It can be shown \citep{Jaakkola:94} that this policy update always increases the average reward, unless the condition
\begin{equation} \label{eq:cond}
\max_a[Q(s,a)-V(s)]>0
\end{equation}
 is satisfied. The algorithm increases the average reward until the condition \eqref{eq:cond} is false, which constitutes a local maximum.

The Jaakkola algorithm consists of two hyper-parameters: the discount factor $\gamma$ and the update rate $\epsilon$. For convergence guarantees, $\gamma$ should initially be selected as a number between zero and one, and should be scheduled in such a way that it converges to 1 in the limit. $\epsilon$, on the other hand, should satisfy $0 \leq \epsilon \leq 1$.

\section{Bringing the pieces together: An interplay between game theory and reinforcement learning}
\label{sec:bring}

\begin{figure}[bth]
\centering
 \includegraphics[trim = 5mm 30mm 15mm 25mm, clip, width=0.9\textwidth]{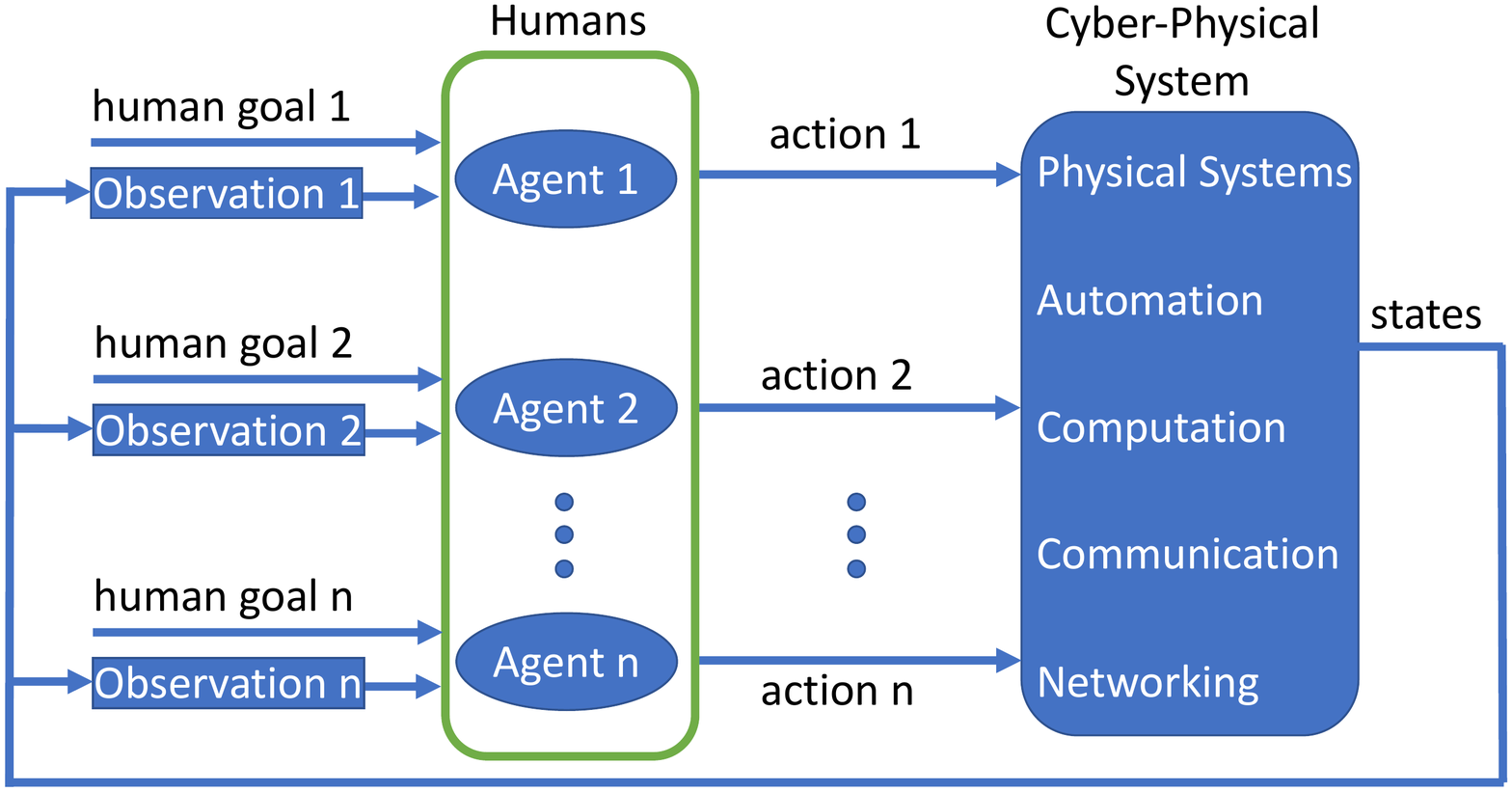}
 \caption{Cyber-physical human system with multiple humans}
 \label{f:CPHS}
\end{figure}

A schematic of the cyber-physical human system (CPHS) we are interested in, involving multiple human interactions, is depicted in Figure~\ref{f:CPHS}. This is a simplified diagram of the overall system, showing how humans' actions change the cyber-physical system states, which are observed by humans who produce the next set of actions and initiate the next cycle accordingly. It is noted that humans' actions are also affecting each other through the closed loop nature of the information flow. Another important take-away from this figure is that human observations are not necessarily the same and an individual human agent does not have full state information. \textit{Observation i} blocks, where $i=1, 2, ..., n$, represent the human observation process. This process is limited and imperfect, therefore each human receives a noisy subset of the whole system state. \textit{Human goal} blocks represent what each of the human agent is trying to accomplish, which can be driving from point A to point B in a road traffic scenario, or protecting his or her own aircraft from a hacker attack in a cyber-security scenario. 

\begin{figure}[bth]
\centering
 \includegraphics[trim = 5mm 30mm 5mm 35mm, clip, width=0.9\textwidth]{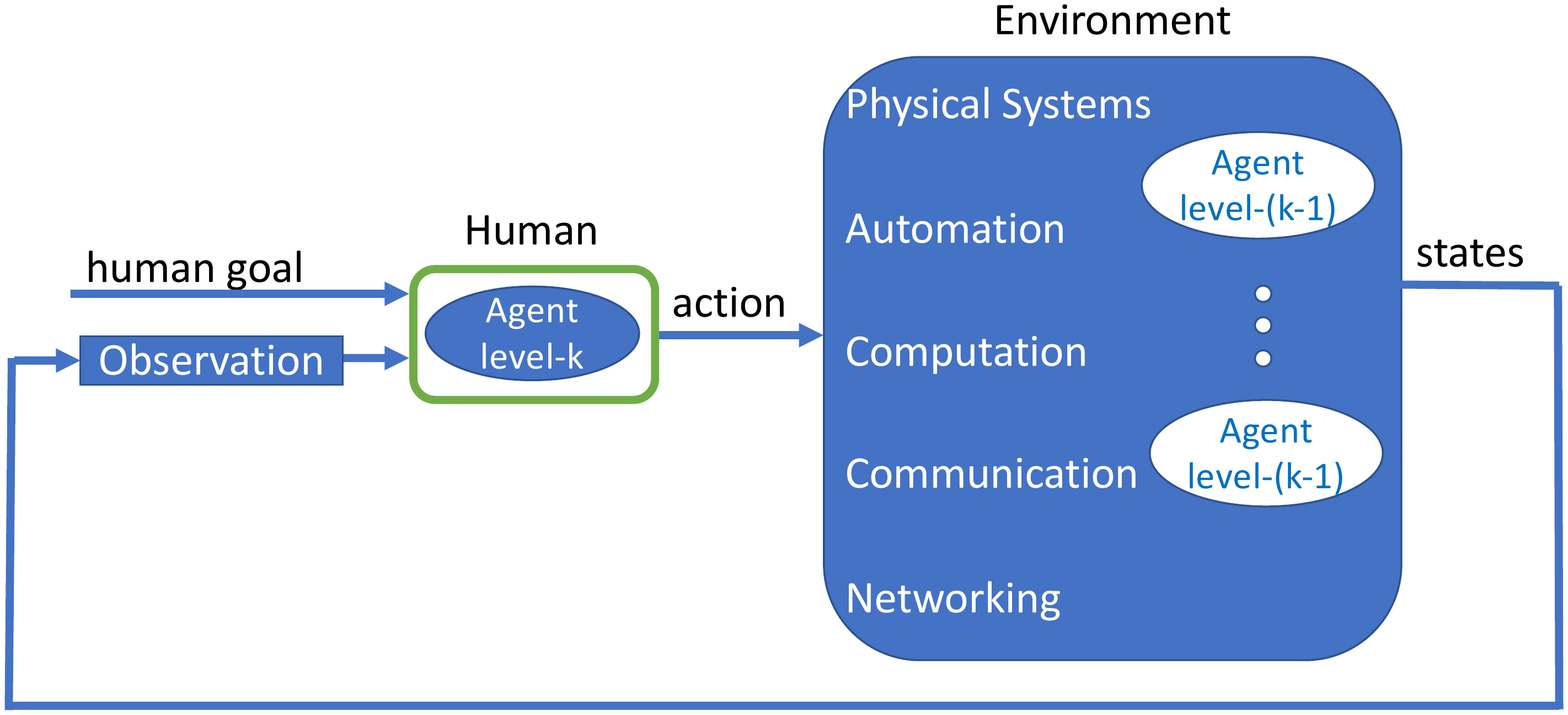}
 \caption{Obtaining level-k policy}
 \label{f:levelk}
\end{figure}

Obtaining a model for the system shown in Figure~\ref{f:CPHS} requires considering the interaction of human actions. Humans are strategic thinkers and therefore if we want to build a model that represents reality, to the best of our ability, we need to find a way to have agent models that consider other agents' possible actions before making a move. As discussed in Section~{\ref{sec:game}}, this is what game theory is all about: Modeling the interactions between strategic agents. Therefore, we may utilize game theory to solve this problem. However, there is one complication in this approach. This is a system where the interaction may last for long periods of time, and therefore obtaining an equilibrium solution can be computationally intractable. If the number of agents get larger, as in several real world applications, even for short periods of interactions, the computational cost grows rapidly. One way around this problem is using the non-equilibrium game theoretical solution concept, level-k thinking, explained in Section~\ref{sec:game}, where human actions are predicted in an iterated manner, instead of being evaluated at the same time. This means that once the \textit{anchoring level}, level-0, is selected, a level-1 human agent's behavior can be identified as the best response to all the other human actions that are determined by the level-0 policy. Similarly, once level-1 behavior is identified, all the agents in the system are assigned the level-1 policy except the one whose level-2 behavior is to be found. Therefore, to predict the policy of a level-k agent, all the rest of the agents' policies are set to level-(k-1), which effectively make them a part of the environment whose dynamics are known, and level-k policy is determined as the best response to the rest of the level-(k-1) policy actions (see Figure~\ref{f:levelk}). This isolates the level-k policy as the single policy that needs to be computed. It is noted that in Figure~\ref{f:levelk}, the actions of the level-(k-1) agents are not shown and the \textit{states} still belong to the CPS. These details are omitted for brevity.  

Once the difficulty of high computational cost due to multiple decision makers is solved using level-k thinking, the problem reduces to estimating the optimal action sequence of an intelligent agent in a given environment.  This problem can be stated as a reinforcement learning (RL) problem by properly defining the states, actions, the environment and the reward function. (Compare Figure~\ref{f:RL} and Figure~\ref{f:levelk}.) Therefore, using one of the suitable RL methods described in Section \ref{sec:RL}, modeling of the CPHS with multiple human interactions can be completed. The type of the preferred RL algorithm depends heavily on the observation and action spaces of the agents, which in turn depend on the engineering domain of the CPHS operation. The general algorithm used in obtaining the agent policies, demonstrating the interplay between the game theory and RL, is provided in Algorithm \ref{a:RLGT}, where $k$ is the maximum desired level. It is noted that, a level-k agent can be made to best-respond to all lower levels, k=0, 1, .. (k-1), instead of only best-responding to level-(k-1), which may or may not be desired depending on the application, by including all lower levels in the agent's policy space.

\begin{algorithm}
	\caption{Interplay between RL and Game Theory}
	\label{alg:RL_GT}
	\begin{algorithmic}[1]
		\STATE Set i = 0
		\WHILE {i $ < $ k}
		\STATE Load the level-i policy
		\STATE Set cognition levels of all players in the environment other than the learning agent to level-i, i.e. set policies of players to level-i policy
		\STATE Place the learning agent in the initialized environment, in which all players are level-i
		\STATE Start the training of the learning agent using a reinforcement learning method, through which agent learns how to best respond to level-i players
		\STATE Once the training is completed, learning agent becomes a level-(i+1) player
		\STATE Save the policy of the learning agent as level-(i+1) policy
		\STATE i += 1
		\ENDWHILE
	\end{algorithmic} \label{a:RLGT}
\end{algorithm}

In the following sections, based on \cite{MusaviJGCD:16} and \cite{Albaba:09}, we explain how this interplay between RL and game theory is used to create modeling frameworks for two different engineering realms, where multiple (180 in one case and 125 in the other) human interactions are involved. 

\begin{remark}
The methods discussed in this paper, which are used to model cyber-physical human systems, are used solely for modeling purposes. Therefore, they are not meant to be used in physical motion systems. 
\end{remark}

\begin{remark}
There exist several successful cyber-physical system models in the literature. The bottleneck in driving models for cyber-physical human systems, is computing the multi-agent, multi-move decision making dynamics, which corresponds to obtaining the models in the green ``Humans" block in Figure~\ref{f:CPHS}. Therefore, in this paper, to emphasize the power of the discussed game theoretical modeling approach in predicting human responses, we avoided employing complicated vehicle models.  
\end{remark}

\begin{remark}
The cyber-physical human systems modeling method explained in this paper merges reinforcement learning and game theory. It is noted that the level-k approach is not the only game theoretical method that can be employed here. Theoretically, other game theoretical methods can also be used. However, as explained in this section, thanks to the hierarchical modeling approach inherent in the level-k reasoning, modeling multi-move, multi-agent scenarios with simultaneous decision makers can be handled in a computationally tractable manner, which makes the level-k method a suitable candidate for the scenarios investigated in this paper. 
\end{remark}

\section{Hybrid airspace modeling}
\label{sec:air}

With \textit{hybrid airspace} we refer to an airspace where manned and unmanned aircraft coexist. As discussed at length in Section \ref{sec:unmanned},  obtaining hybrid airspace models is a necessity for successful integration of unmanned aircraft systems (UAS) into the National Airspace System (NAS). In this section, we present how the game theoretic modeling framework discussed here is used to realize the modeling of these systems. 

\begin{figure}[bth]
\centering
 \includegraphics[trim = 0mm 90mm 0mm 95mm, clip, width=0.9\textwidth]{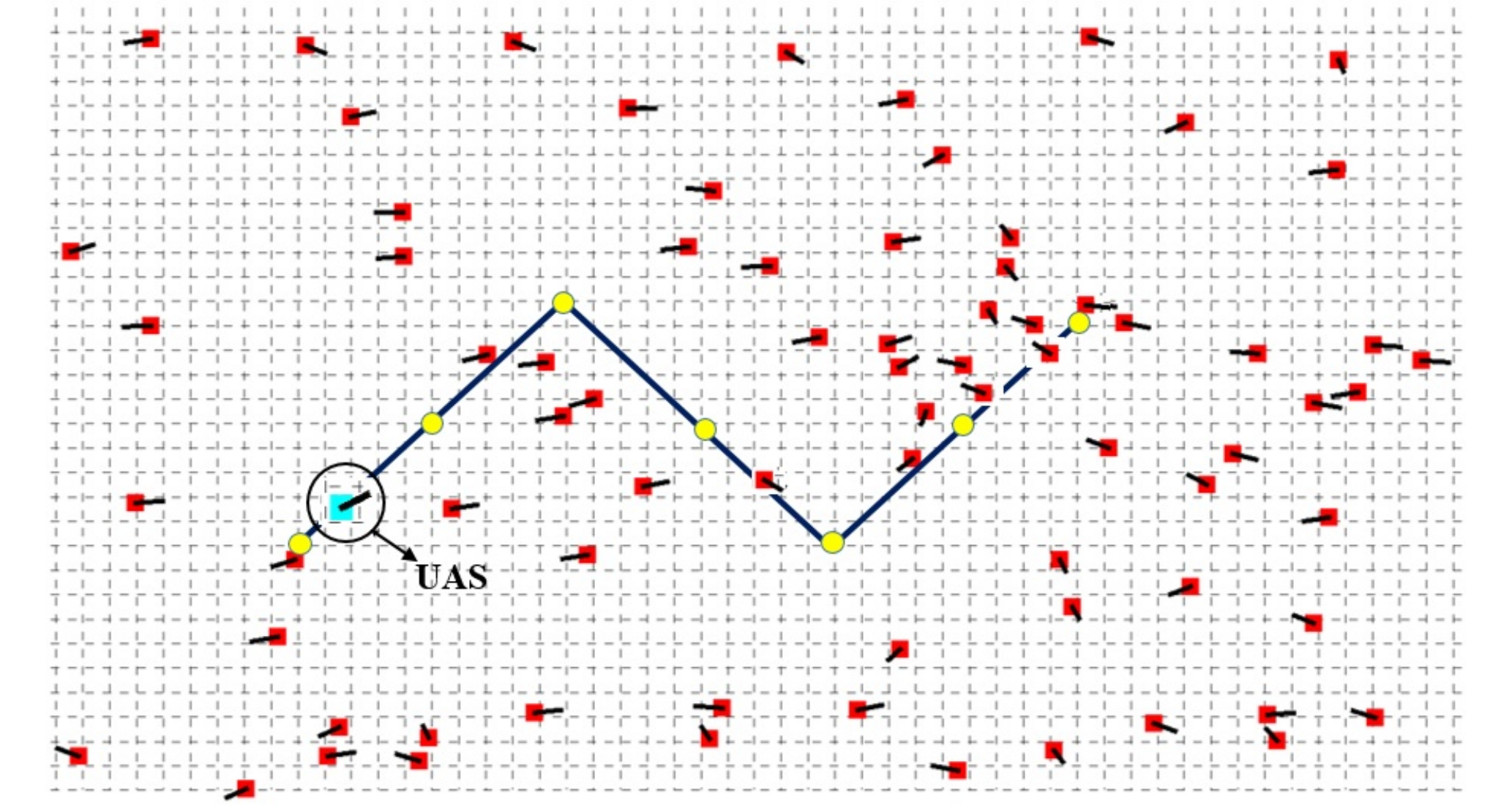}
 \caption{A hybrid airspace scenario where red squares represent manned aircraft and the cyan square is used to indicate an unmanned aircraft (UA). The yellow circles in the 600 km x 300 km airspace are the waypoints assigned to the UA. (\cite{MusaviJGCD:16}, reprinted with permission of the American Institute of Aeronautics and Astronautics, Inc.)}
 \label{f:CrowdedScenarioUAS}
\end{figure}

Figure \ref{f:CrowdedScenarioUAS} shows a hybrid airspace scenario where a UAS (cyan) is assigned to follow certain waypoints (yellow) in a crowded airspace filled with manned aircraft (red). We are interested in how the overall system will evolve in time. 

To be able to predict the possible outcomes of this scenario, we need to create a model that will capture the reaction dynamics of manned aircraft pilots in cases of separation conflicts. As discussed earlier, the game theoretical modeling approach produces policies, which are probabilistic maps from observations to actions, to represent human reactions. Therefore, to obtain these policies, we first need to clearly define the \textit{observation} and \textit{action spaces} and represent them in a way that is meaningful for the RL algorithm. Once these spaces are explicitly defined, pilot goals and preferences need to be expressed in form of a \textit{reward function}. Furthermore, to train pilot policies in a simulation environment, we need to realize the motions of the aircraft, manned and unmanned, using their \textit{physical models}. Finally, \textit{sense and avoid algorithms} need to be integrated to UAS dynamics for collision avoidance. Below, we explain how these pieces are obtained and then assembled to create the overall hybrid airspace model. Furthermore, we discuss the validation studies conducted using data.

\subsection{Observation and action spaces} 
\label{sec:obs}

\begin{figure}[bth]
\centering
 \includegraphics[trim = 0mm 0mm 0mm 0mm, clip, width=0.9\textwidth]{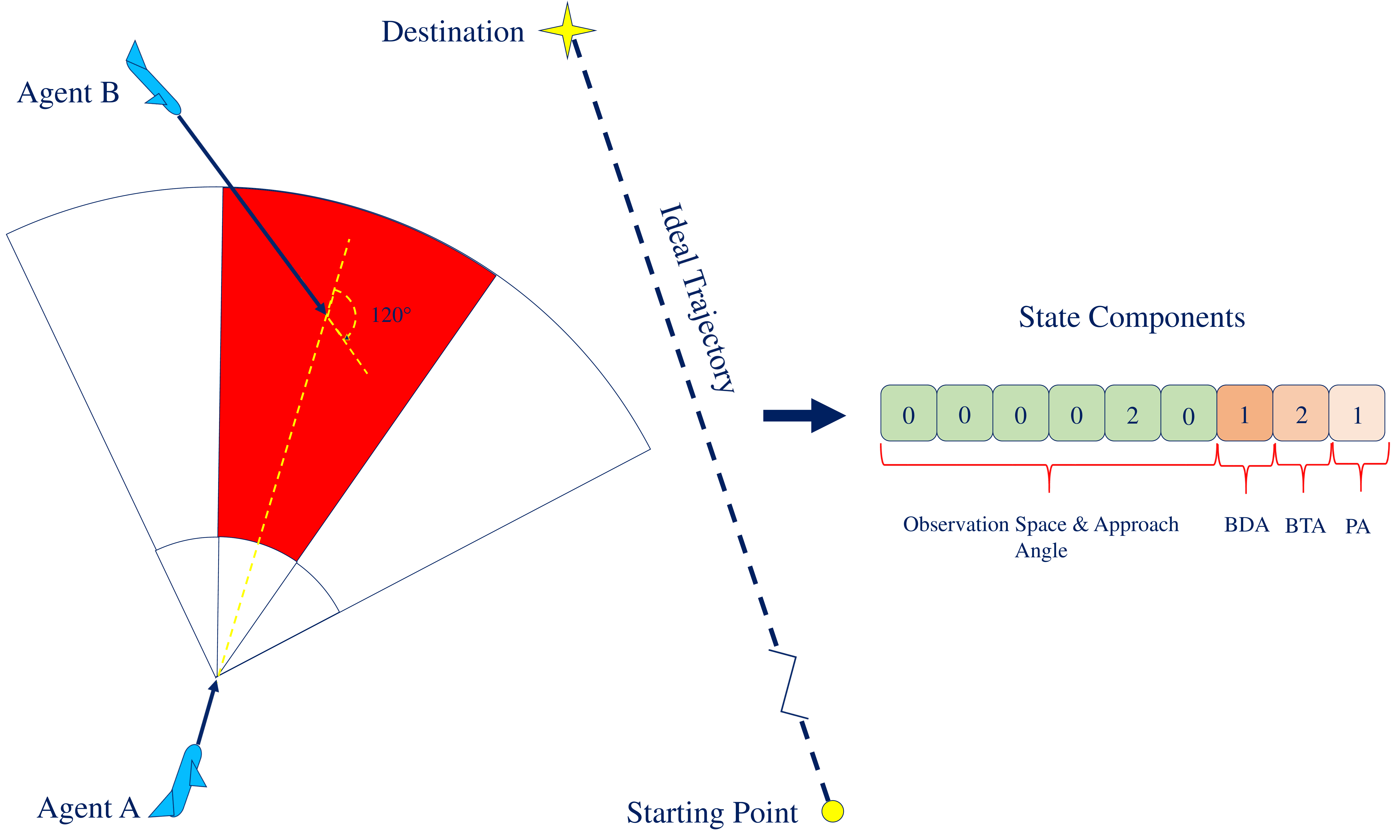}
 \caption{Observation space. (\cite{MusaviJGCD:16}, reprinted with permission of the American Institute of Aeronautics and Astronautics, Inc.)}
 \label{f:Obs}
\end{figure}

Self-separation concept, where the pilots (and crew) are responsible for keeping a safe distance from encountered traffic, is a procedure that is being explored for Next Generation (NextGen) Air Transportation System \citep{wing2013pilot}. One of the technologies that can make this possible is Automatic Dependent Surveillance Broadcast (ADSB), which provides state information of the surrounding traffic, with a precision better than the radar \citep{kacem2018ads}. In the above scenario, we assume that aircraft are equipped with this technology. To factor in cognitive limitations, the observation space is set as a pie shaped region formed by two circles sharing a common center, which is depicted in Figure~\ref{f:Obs}, on the left. The inner circle radius is taken as 1~nmi and the outer circle radius is taken as 5~nmi, reflecting the standard separation requirements for manned aircraft \citep{Batlle:12}. These circles are then divided into 3 slices resulting in 6 different observation regions. If an intruder approaches one of these regions, the region that is being approached is coded with number 1, 2, 3 or 4, depending on the approach angle, while the rest obtains a zero value. For example, in Figure~\ref{f:Obs}, Aircraft B  is approaching Aircraft A's observation space with a 120 degree angle which makes the corresponding entry for the approached region take the value of 2, while the rest are assigned 0. This is shown under the \textit{state components} section on the rightmost side of the figure. The information of the best immediate actions that will make the aircraft approach its predetermined trajectory (Best Trajectory Action, BTA) and to its predetermined destination (Best Destination Action, BDA), together with the pilot's Previous Action (PA) are also coded into the state components. The pilot action space consists of 3 actions: \textit{45 degrees left}, \textit{straight} and \textit{45 degrees right}, and these actions are coded as 0, 1 and 2, respectively, in the state components. It is noted that these actions go through aircraft dynamics to produce the actual aircraft motion. During training, the state components entries are continuously updated, based on the explanations given above, and used in the RL algorithm. With the aforementioned observation and action spaces, the RL algorithm need to assign Q values to $5^6\times 3^3 \times 3 = 1,265,625 $ state-action pairs.
\begin{remark}
The actions described in this section are not the actions of the aircraft but the decision inputs to the real continuous aircraft dynamics, whose outputs are the aircraft actions. The aircraft dynamics is explained in Section~\ref{sec:physical}.
\end{remark}

\subsection{Reward function} 
\label{sec:reward}

For the RL algorithm to evaluate the desirability of selecting a certain action given a state, a reward function that specifies pilot preferences and goals is required. For the UAS integration into NAS scenario, the reward function 
\begin{equation} \label{eq:rewardNAS}
r = -\omega_1 C -\omega_2 S -\omega_3 I +\omega_4 D + \omega_5 P - \omega_6 E
\end{equation}
is used, where $C$ and $S$ represent the number of aircraft (manned or unmanned) occupying a space within the collision and separation regions of the ego aircraft, respectively. The definition of these regions can be found at \cite{planning2007concept} and \cite{Batlle:12}. $I$ is binary and gets the values of 1 or 0 depending on whether the ego aircraft is approaching or distancing from the intruder. $D$ and $P$ are the degrees of approach (or distancing) by the ego aircraft to its destination and to its defined trajectory, respectively, normalized by the distance covered in one time step. To accommodate the tendency of humans to minimize energy consumption as much as possible, $E$ is introduced to represent \textit{effort}, taking the values of 1 or 0, depending on whether the pilot takes a new action or not, respectively.

\begin{remark}
The selection of the reward function plays a crucial role in the performance of the reinforcement learning algorithm. A formal procedure for the determination of a proper reward function is an open area of research. However, addressing context-dependent problems that reflect the agent\textquotesingle s preferences, while keeping the function simple is the general approach for determining the reward functions. One problem that needs to be solved in this section is avoiding collisions and separation violations, which is addressed by the terms $C$, $S$ and $I$ in \eqref{eq:rewardNAS}. Another problem is reaching a given destination while following a predetermined trajectory, which is addressed by the terms $D$ and $P$. Finally, the problem of energy conservation is addressed by the term $E$. An alternative to \eqref{eq:rewardNAS} could be a reward function that uses continuous variables for all the terms. This would increase the resolution but could unnecessarily complicate the function. In certain problems such as learning a policy from an expert, the reward function selection can be done using a more systematic approach by solving an inverse reinforcement learning problem \citep{sutton2018reinforcement}.
\end{remark}

\subsection{Physical models of manned and unmanned aircraft} 
\label{sec:physical}

As explained above, aircraft are controlled by pilots' commands of heading angle changes ($\pm45$ degrees). Using the standard turn rate of 3 deg/s angular velocity \citep{nancy2016psilot}, we model the aircraft turning motion with a first order dynamics having a time constant of 10 seconds. The related differential equation is
\begin{equation}
\dot{\Psi}=-0.1(\Psi-\Psi_d),
\end{equation}
where $\Psi$ and $\Psi_d$ represent the current and the desired heading angles, respectively. The manned aircraft are assumed to be flying in en route phase with a constant velocity $v$, having the $x$ and $y$ coordinate components
\begin{equation}
v_x=|v|\sin \Psi
\end{equation}
and
\begin{equation}
v_y=|v|\cos \Psi.
\end{equation}
Unmanned aircraft are assumed to fly autonomously while avoiding collisions using an onboard sense and avoid (SAA) algorithm. This algorithm commands velocity vector changes if an intruder is detected. Using a 1 second time constant \citep{mujumdar2011reactive} and a first order dynamics, the velocity vector dynamics are modeled as
\begin{equation}
\dot{\vec v}=-(\vec v-\vec v_d), 
\end{equation}
where $\vec v_d$ is the desired velocity vector. Two SAA algorithms are employed in the simulation environment with different velocity vectoring properties. Both of them first detects a probable conflict by forecasting the future trajectories of both the ego and intruder aircraft, and checking whether the minimum calculated distance, $R$, between them is smaller than a predetermined threshold value. If a conflict is detected, a desired velocity vector command, $\vec v_d$, is produced. One of the SAA algorithms (SAA1) proposed by \cite{fasano2008multi} issues the velocity command 
\begin{equation}
{\vec v_d}=\Bigg( \frac{\vec v_{ei} \cos(\eta-\xi)}{\sin(\xi)}  \bigg( \sin(\eta) \frac{\vec v_{ei}} {|\vec v_{ei}|} - \sin(\eta - \xi) \frac{\vec r}{|\vec r|} \bigg) + \vec v_i  \Bigg),
\end{equation}
in cases of conflict, where the ego unmanned aircraft velocity command and intruder aircraft velocity are represented by ${\vec v_d}$ and $\vec v_i$, respectively. Similarly, $\vec r$ and $\vec v_{ei}$ refer, respectively, to the relative position and velocity between these two aircraft, and $\eta$ is the angle between these two vectors. Angle $\xi$ is determined as $\xi=\sin^{-1}(R/|\vec r|)$. The other SAA algorithm (SAA2) used in simulations provides the velocity command 
\begin{equation}
\label{e:function}
{\vec v_d}=
\frac{-\vec{v}_{e}(\frac{\vec{r}_{0} \cdot  \vec{v}_{ei}}{|\vec{v}_{ei}|})-(R-|\vec{r}_{m}|)\frac{\vec{r}_{m}}{|\vec{r}_{m}|}}{\bigg|-\vec{v}_{e}(\frac{\vec{r}_{0} \cdot \vec{v}_{ei}}{|\vec{v}_{ei}|})-(R-|\vec{r}_{m}|)\frac{\vec{r}_{m}}{|\vec{r}_{m}|}\bigg|}
\end{equation}
to resolve conflicts, where $\vec{r}_{0}$ and $\vec{r}_{m}$ stand for the initial and minimum relative positions between the ego unmanned aircraft and the intruder \citep{mujumdar2011reactive}.

\subsection{Simulation results} 
\label{sec:sim}
After the necessary pieces explained above are set to create the overall hybrid airspace model, pilot reactions are obtained using the interplay, explained in Section \ref{sec:bring}, between the level-k game theoretical approach (Section \ref{sec:game}) and Jaakkola reinforcement learning method (Section \ref{sec:Jaak}). \textit{The algorithm that details this process is given in Algorithm \ref{alg:NAS}.} Once the pilot reaction dynamics are created, quantitative analyses on the integration scenario are conducted using Monte Carlo simulations. 
\textit{\begin{algorithm}
	\caption{Interplay between RL and Game Theory in NAS}
	\label{alg:RL_GT_NAS}
	\begin{algorithmic}[1]
		\STATE $i = 0$
		\WHILE {$i  < k$ ($k$ is the maximum cognition level)}
		\STATE Load the level-i policy
		\STATE Set the policies of all the pilots in the scenario, other than the ego pilot (the pilot being trained), to level-i
		\STATE Start the training of the ego pilot using reinforcement learning, through which the pilot learns how to best respond to level-i pilots
		\STATE Once the training is completed, the ego pilot becomes a level-$(i+1)$ pilot
		\STATE Save the policy of the ego pilot as level-$(i+1)$ policy
		\STATE $i += 1$
		\ENDWHILE
	\end{algorithmic} \label{a:RLGTNas}
	\label{alg:NAS}
\end{algorithm}
}
Since one of the bottlenecks of UAS integration into NAS is the maturation of SAA algorithms, we present how the modeling method discussed in this article can be used to conduct comparative quantitative analysis on various aspects of different SAA methods. Specifically, we analyze the effect of \textit{distance horizon} and \textit{time horizon} variables used in the SAA development. In this article, we define the former as the scan radius of the algorithm, and the latter as the amount of projection time used to detect a conflict. 

\begin{figure}[!htb]	
	\centering
	\begin{subfigure}[htb]{10cm}
			\centering
			\includegraphics[width=10cm,height=5cm,keepaspectratio]{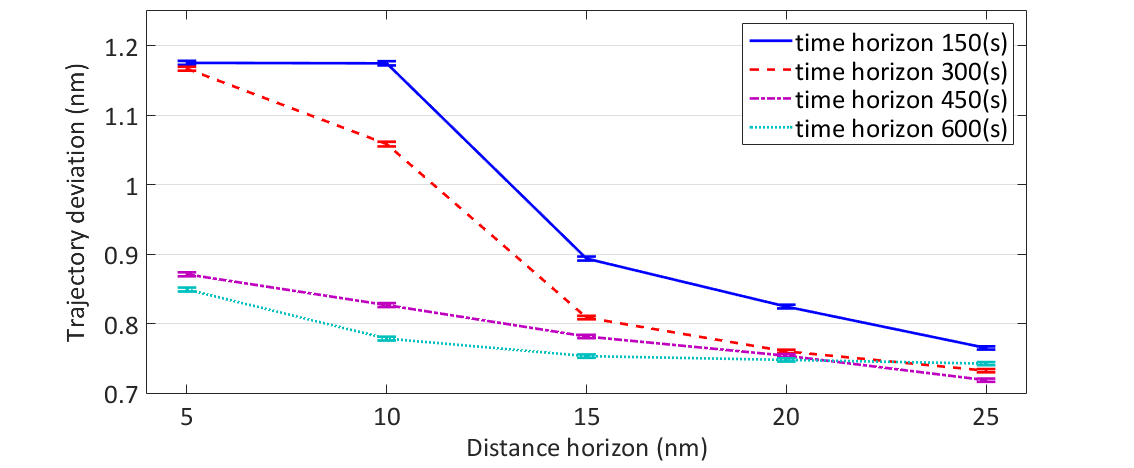}
		\caption{manned aircraft trajectory deviation}\label{f:SAA1_1}		
	\end{subfigure}
	\quad
	\begin{subfigure}[htb]{10cm}
			\centering
			\includegraphics[width=10cm,height=5cm,keepaspectratio]{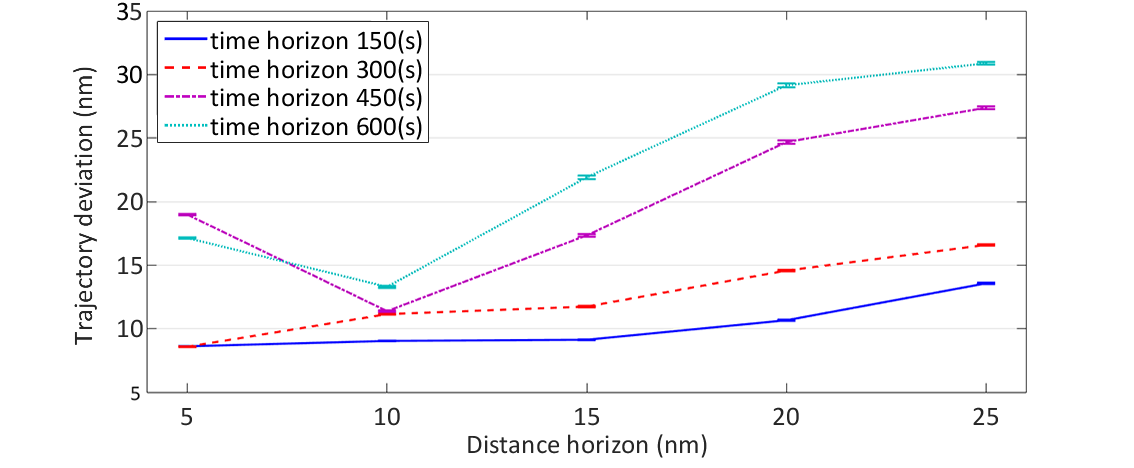}
		\caption{UAS trajectory deviation}\label{f:SAA1_2}
	\end{subfigure}
	\quad
	\begin{subfigure}[htb]{10cm}
			\centering
			\includegraphics[width=10cm,height=5cm,keepaspectratio]{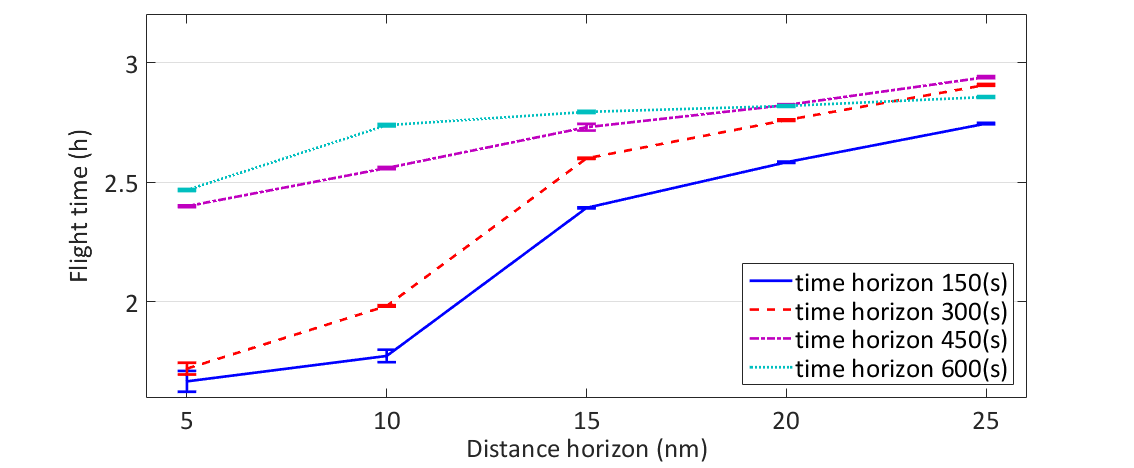}
			\caption{UAS flight time}\label{f:SAA1_3}
	\end{subfigure}
	\quad
	\begin{subfigure}[htb]{10cm}
			\centering
			\includegraphics[width=10cm,height=5cm,keepaspectratio]{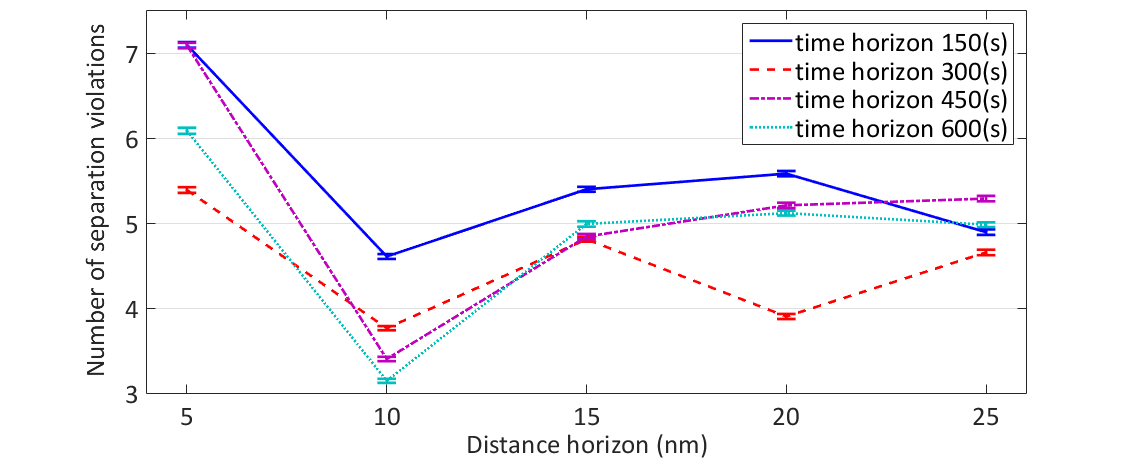}
			\caption{number of separation violations}\label{f:SAA1_4}
	\end{subfigure}
	\caption{When SAA1 is employed in HAS, effects of varying time and distance horizons on separation violations, flight times and trajectory deviations are presented. (\cite{MusaviJGCD:16}, reprinted with permission of the American Institute of Aeronautics and Astronautics, Inc.)}\label{f:SAA1}
\end{figure}

\begin{figure}[H]	
	\centering
	\begin{subfigure}[h]{10cm}.
			\centering
			\includegraphics[width=10cm,height=5cm,keepaspectratio]{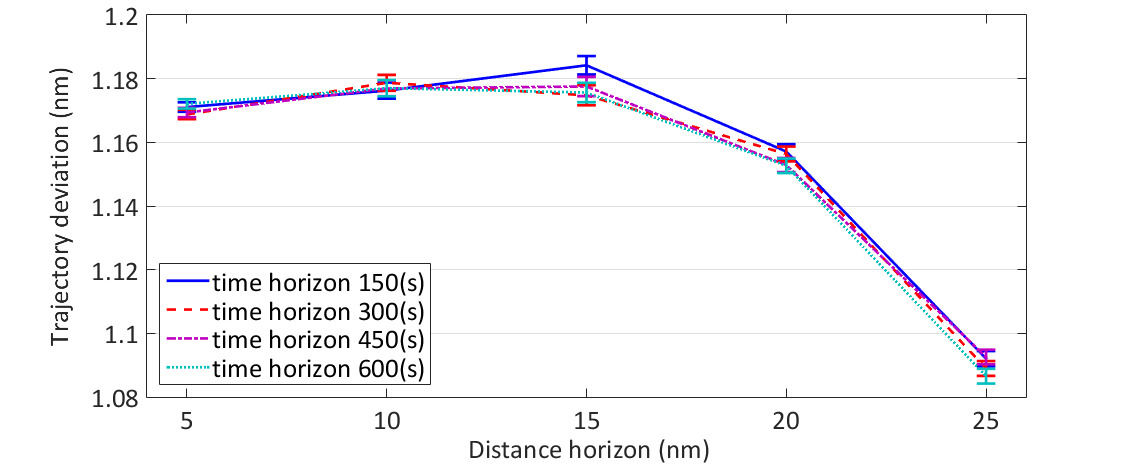}
		\caption{manned aircraft trajectory deviation}\label{f:SAA2_1}		
	\end{subfigure}
	\quad
	\begin{subfigure}[h]{10cm}
			\centering
			\includegraphics[width=10cm,height=5cm,keepaspectratio]{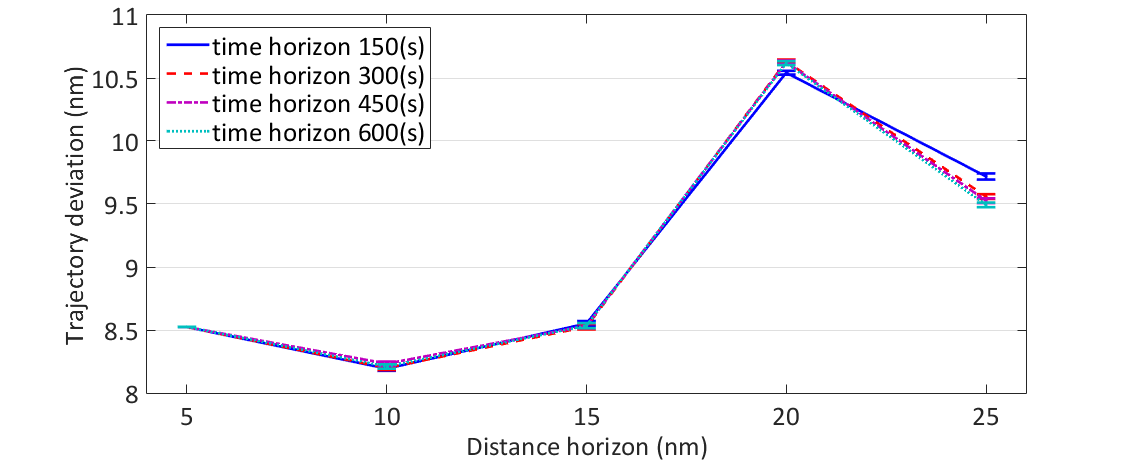}
		\caption{UAS trajectory deviation}\label{f:SAA2_2}
	\end{subfigure}
	\quad
	\begin{subfigure}[h]{10cm}
			\centering
			\includegraphics[width=10cm,height=5cm,keepaspectratio]{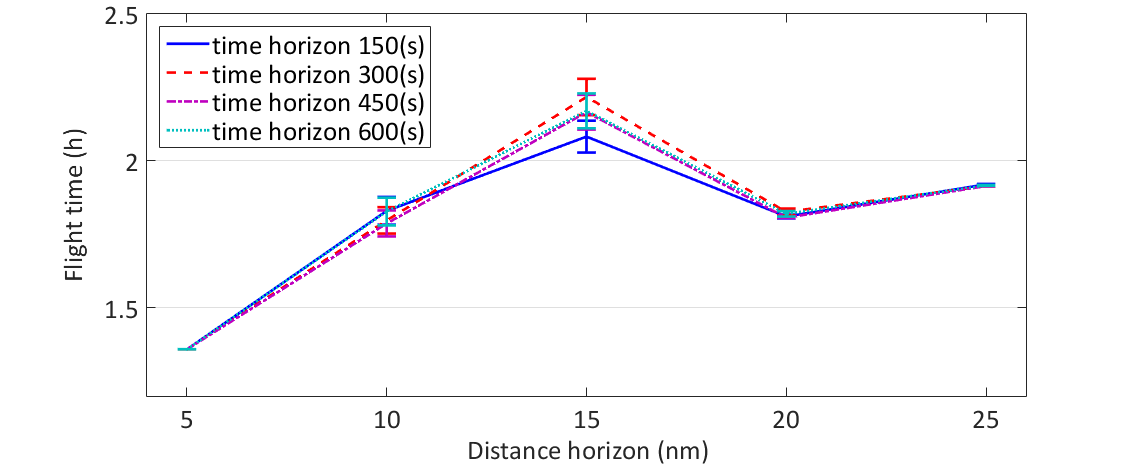}
			\caption{UAS flight time}\label{f:SAA2_3}
	\end{subfigure}
	\quad
	\begin{subfigure}[h]{10cm}
			\centering
			\includegraphics[width=10cm,height=5cm,keepaspectratio]{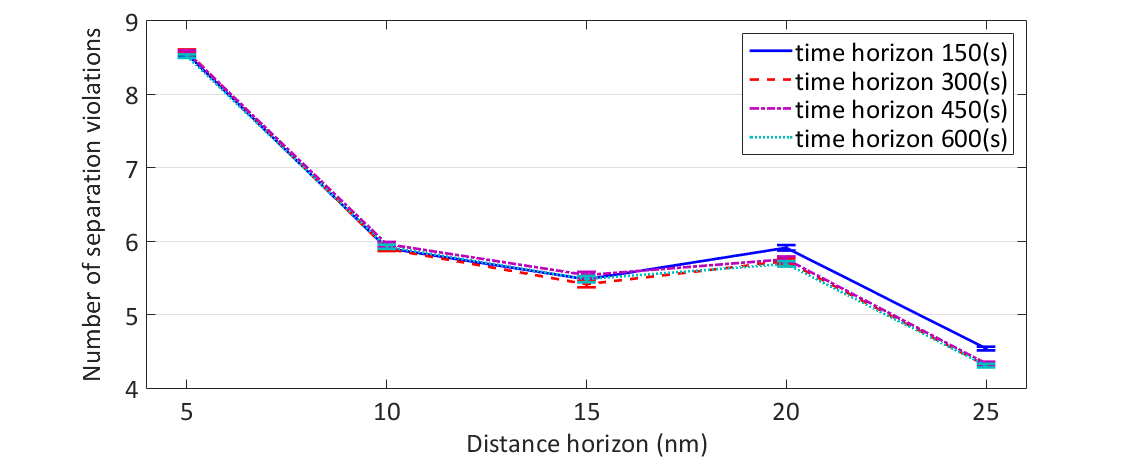}
			\caption{number of separation violations}\label{f:SAA2_4}
	\end{subfigure}
	\caption{Safety vs. performance in HAS, when SAA2 is employed is depicted by showing the changes in trajectory deviations, flight times and separation violations with the changes in distance and time horizons. (\cite{MusaviJGCD:16}, reprinted with permission of the American Institute of Aeronautics and Astronautics, Inc.)}\label{f:SAA2}
\end{figure}

Figure~\ref{f:SAA1} and Figure~\ref{f:SAA2} depict the effect of varying distance and time horizons on trajectory deviations, flight times and number of separation violations, for SAA1 and SAA2, respectively. A few conclusions, not in increasing or decreasing importance, can be drawn from these results. First, although time horizon makes a significant effect on both safety (separation violations) and performance (trajectory deviations and flight times) measures for SAA1, the unmanned aircraft equipped with SAA2 is not affected as much with the variation of this parameter. Second, the effect of the distance horizon on separation violation numbers levels out quickly for SAA1, while SAA2 keeps showing lesser and lesser violation rates, although SAA1 provides a safer traffic in the same parameter range. Third, UAS trajectory deviations are generally smaller when SAA2 is employed. Fourth, an interesting but expected phenomenon is observed: the trajectory deviations of manned and unmanned aircraft have negative correlation. As the UAS trajectory deviations increase, showing that SAA is doing more work for keeping a safe distance, manned aircraft pilots spend less effort and have less trajectory deviations. All these results obtained in a simulation environment with real-time decision making pilots can be used in testing and tuning of SAA algorithms, as well as making a quantitative comparison between different approaches. 

\subsection{Validation} 
\label{sec:sim}
Although we do not have enough, if at all, UAS integration data to test against the predictions of the game theoretical modeling approach, the resulting models created by the method need to be validated using other means, to achieve at least a minimum level of credibility. Several validation methods are introduced in the literature, such as \textit{face validity}, \textit{comparison with validated models}, \textit{historical data validation}, \textit{parameter sensitivity analysis} and \textit{predictive validation} \citep{verify}. Among these, the ones that are based on data are the most effective validation methods. It is noted that when the data becomes available, one needs to test if relevant statistics between the data and the model are matching. For example, average aircraft trajectory deviation and the average number of separation violations can be compared. Furthermore, during individual encounters, pilot decisions and minimum distances between UAS and manned aircraft, predicted by the model and obtained from the data can be investigated to see whether or not they show similar characteristics. It is noted that the game theoretical modeling approach has enough degrees of freedom, such as reward function terms and weights, to be able to be modified, in case some discrepancies between data and the model are observed when the data becomes available in the future. This is an important feature that needs to be found in predictive models \citep{law2008build}. Since currently data is not available, we discuss two methods that can be used without data: face validation and comparison with a validated model.

Face validation evaluates two aspects of the model: the logic behind the modeling method, and input-output relationships of the model. Two main approaches used in the proposed model, namely reinforcement learning and game theory, are well-established fields that showed promise in modeling real-life behavior. Experimental backing of the level-k game theory is also provided in relevant references discussed in the review of existing work sections in this article. Also, the logic behind the selection of reward function terms is explained in earlier sections. We also discussed above the input-output relationships when we explained the effects of different parameters on the airspace scenario with figures, and showed that the results are reasonable. 

\begin{figure} [htb]	
	\centering
	\begin{subfigure}[htb]{7cm}
		\centering
		\includegraphics[width=7cm,height=5cm,keepaspectratio]{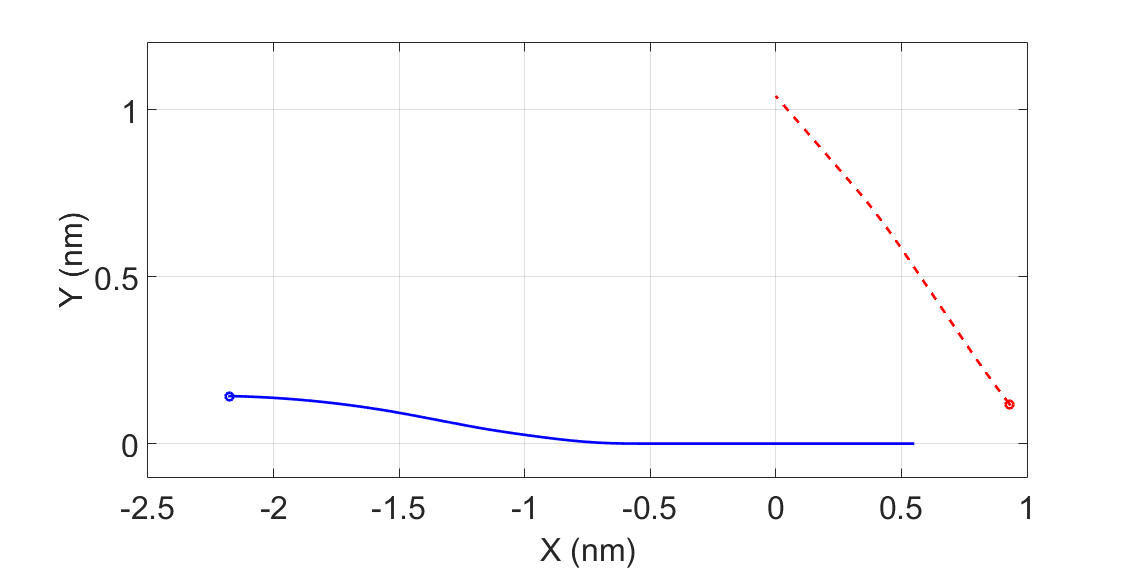}
		\caption{Trajectories created by the validated model.}\label{f:enc3_2}		
	\end{subfigure}
	\quad
	\begin{subfigure}[htb]{7cm}
		\centering
		\includegraphics[width=7cm,height=5cm,keepaspectratio]{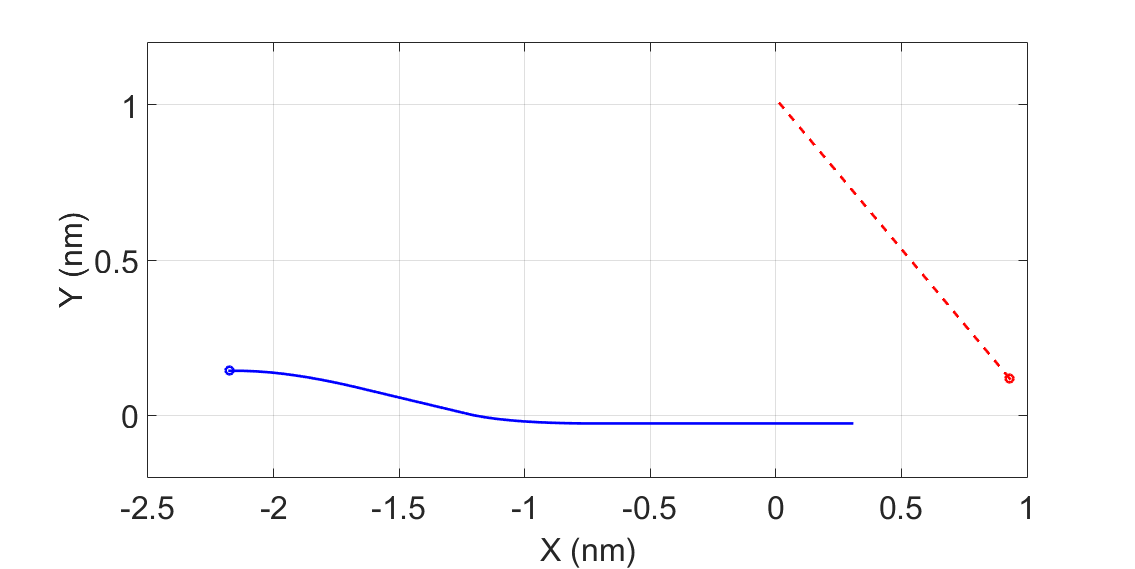}
		\caption{Trajectories created by the proposed model.}\label{f:enc3_1}
	\end{subfigure}
	
	\begin{subfigure}[htb]{7cm}
			\centering
			\includegraphics[width=7cm,height=5cm,keepaspectratio]{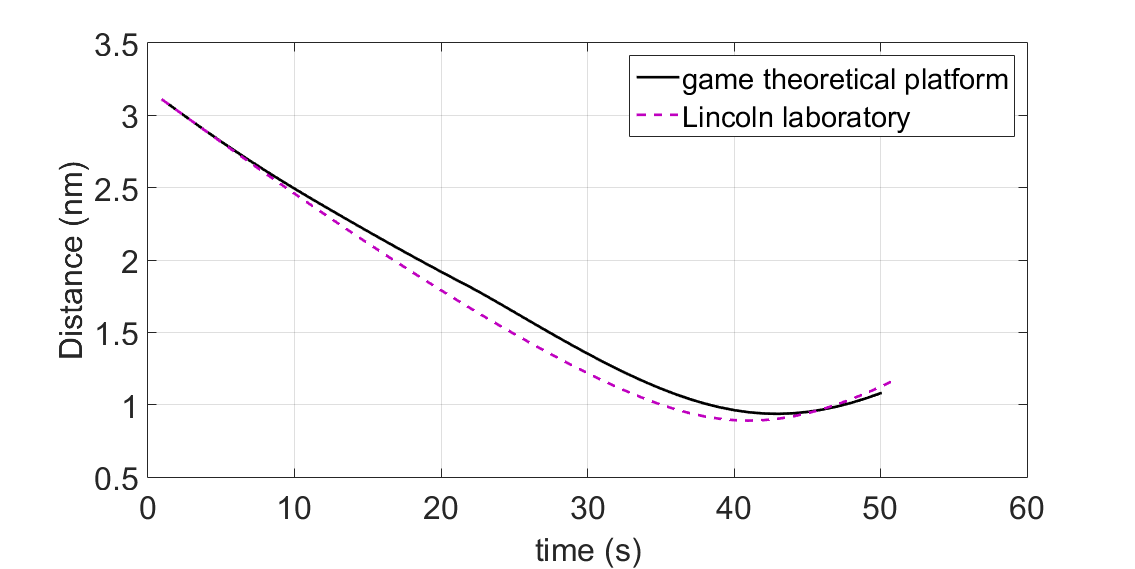}
			\caption{Separation distances for each model.}\label{f:enc3_3}
	\end{subfigure}
	\caption{Comparison of the trajectories created by the validated model and the game theoretical modeling approach for sample encounter number 3. (\cite{MusaviJGCD:16}, reprinted with permission of the American Institute of Aeronautics and Astronautics, Inc.)}\label{f:enc3}
	\vspace{1.1cm}
\end{figure}

Sample trajectories of a data-validated model developed by the Lincoln Laboratory \citep{kochenderfer2008correlated} is available online in two text files that are open to public: \texttt{cor\_ac1.txt} and \texttt{cor\_ac2.txt}. A comparison of one of the available trajectories (encounter 3) with the game theoretical model prediction is provided in Figure \ref{f:enc3}. In the figure, starting points of aircraft are indicated by thick dots. It is seen that pilot decisions, as well as minimum distance between aircraft during the encounter are similar. Further trajectory comparisons between the game theoretical model and this data-validated model can be found in \cite{MusaviJGCD:16}.

\section{Road traffic modeling}
\label{sec:road}

Similar to the unmanned airspace system (UAS) integration study, to obtain the model of the road traffic, we need the physical models of the cars, driver observation and action spaces, and a reward function reflecting the goals and preferences of the drivers. Below, we explain these components and also provide validation of the resultant overall model with traffic data.

\subsection{Driver observation and action spaces}
\begin{figure}[bt]
\centering
\includegraphics[width=0.6\linewidth]{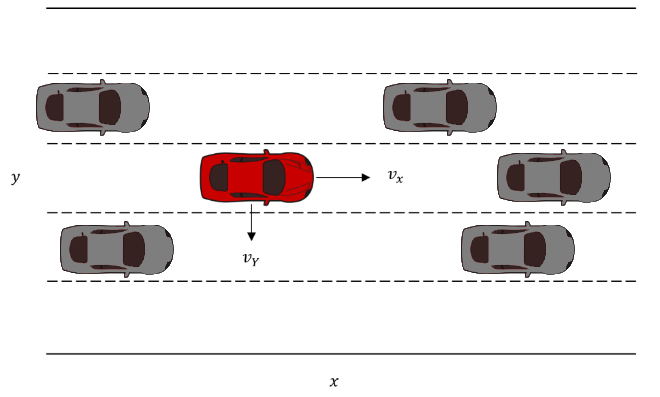}
\caption{Ego vehicle and surrounding traffic}
\label{fig_model}
\end{figure}

\begin{figure}[bt]
		\centering
		\includegraphics[width=0.5\linewidth]{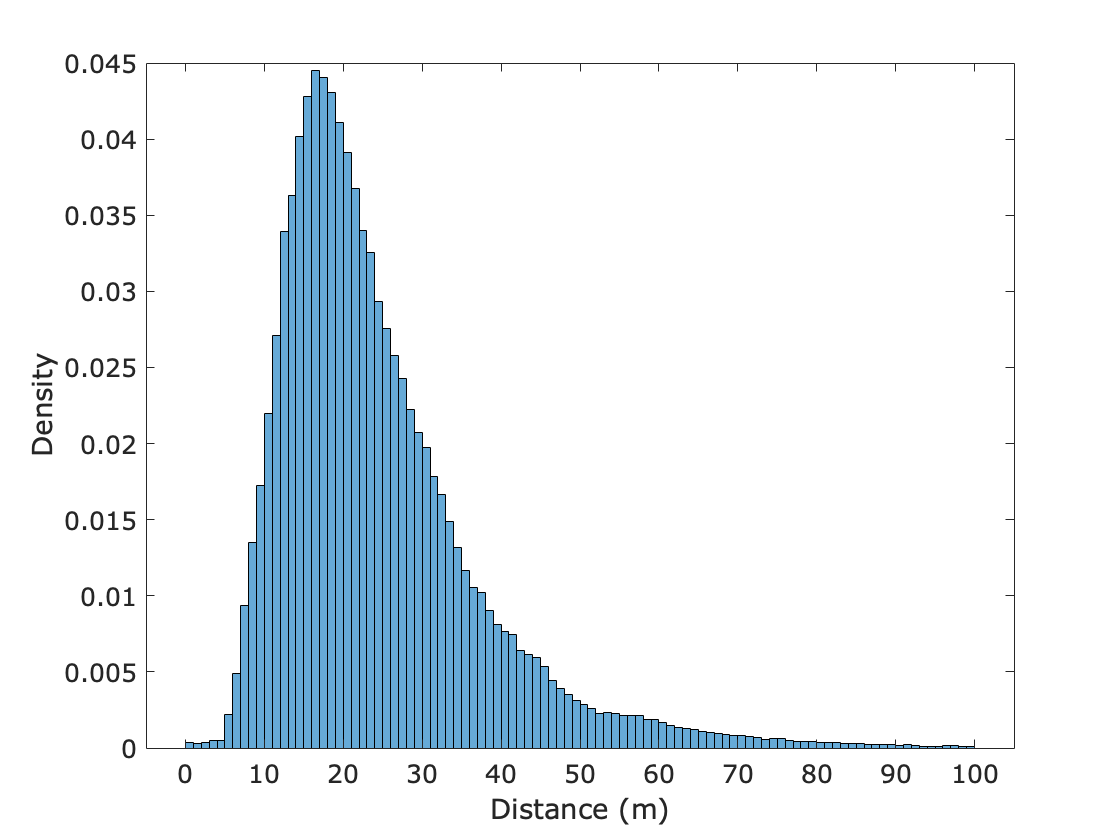}
		\caption{Distribution of distances to car in front.}
		\label{fig_spcdist}
\end{figure}

Several different traffic scenarios can be modeled via the proposed approach. To be able to test the results with data, we developed the model of a 5-lane highway, similar to the US101 Hollywood Freeway, whose raw data is available at \cite{Colyar:07}. In this scenario, the human drivers are assumed to be observing (or being able to process from all available data), his or her immediate neighboring lane cars (front left, front right, rear left, rear right) and the car in front. Figure \ref{fig_model} presents a snapshot of a typical ego vehicle (red) motion, where the driver can observe the surrounding 5 cars. As in the UAS integration scenario, the observation space is quantized and the distances are coded as \textit{nominal}, \textit{close} and \textit{far}. It is noted that this quantization introduces noise and uncertainty to driver observations since instead of the exact location of neighboring cars, only a certain region occupied by the car is known. The importance of introducing noise and uncertainty to the human observations are discussed in Section \ref{sec:bring} and represented by Observation blocks in Figure~\ref{f:CPHS}.
To determine reasonable values for quantization, the distance distribution between cars are processed from the raw data provided in \cite{Colyar:07} and plotted in Figure~\ref{fig_spcdist}. Based on this distribution, \textit{nominal} range is defined to be between 11~m - 27~m, \textit{close} is defined as smaller than 11~m and \textit{far} is used for distances larger than 27~m. It is noted that nominal region consists of approximately half of the area under the curve depicted in the figure. Once the positions of the surrounding cars are defined, their relative motions against the ego car are expressed as \textit{stable}, \textit{approaching} and \textit{distancing}. As a result, the observation space of the ego car consists of quantized positions and relative motions of the surrounding cars.

\begin{figure}[bt]
\centering
\includegraphics[width=0.5\linewidth]{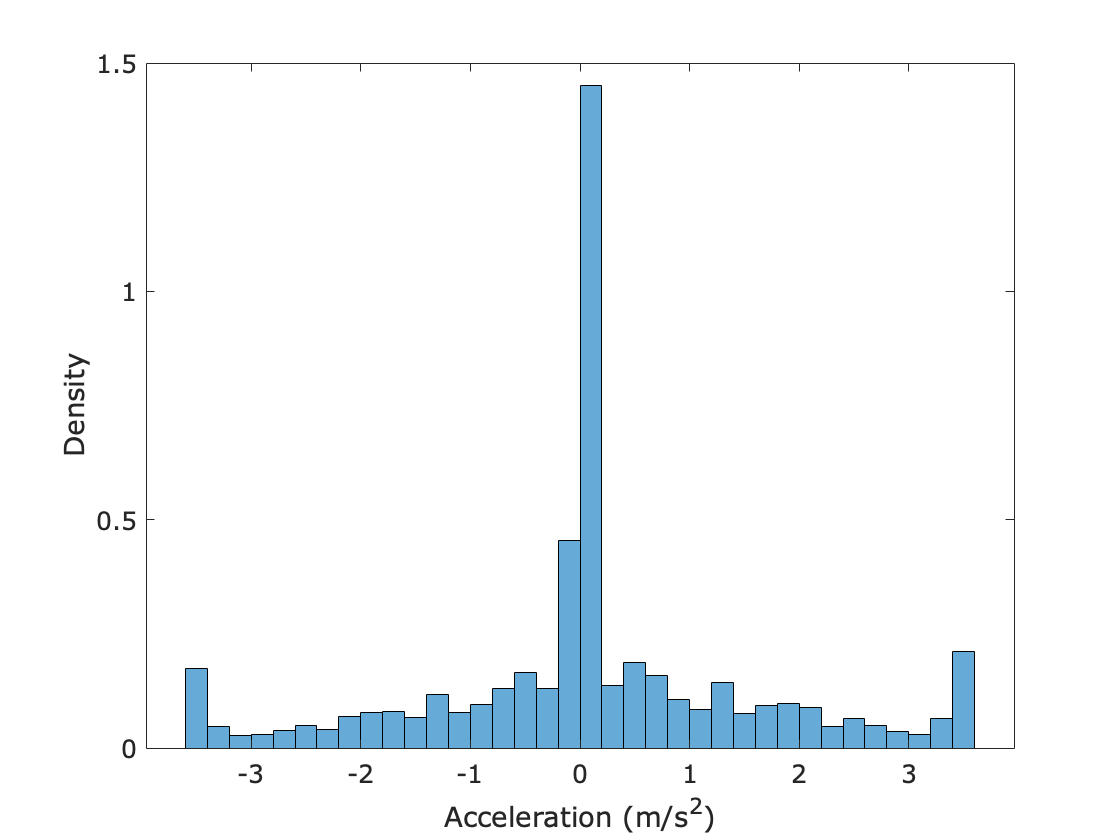}
\caption{Acceleration distribution}
\label{fig_acc}
\end{figure}

As we use traffic data to obtain a meaningful observation space, part of the action space is also formed by considering the acceleration distribution, the plot of which is given in Figure~\ref{fig_acc}, obtained by processing the same traffic data. 
According to this acceleration distribution, acceleration inputs of the drivers are categorized into 5 separate continuous sub-distributions: 1) \textit{Maintain}, a normal distribution with zero mean and 0.075 standard deviation, 2) \textit{accelerate} a uniform distribution between 0.5 m/s$^2$ - 2.5 m/s$^2$, 3) \textit{decelerate} a uniform distribution between -0.5 m/s$^2$ and -2.5 m/s$^2$, 4) \textit{hard accelerate}, a half normal distribution with a 3.5 m/s$^2$ mean and 0.3 m/s$^2$ standard deviation, 5) \textit{hard decelerate}, a half normal distribution with -3.5 m/s$^2$ mean and 0.3 m/s$^2$ standard deviation. Drivers sample these distributions to create an acceleration action, if they choose to. Two other action choices for the drivers are \textit{move to the left lane} and \textit{move to the right lane}, during both of which the velocity is assumed to remain constant. 

\subsection{Reward function}
The driver reward function used in the reinforcement learning (RL) algorithm is 
\begin{equation} \label{eq:rewardTr}
r = \omega_{1}C + \omega_2S + \omega_3D + \omega_4E,
\end{equation}
where $C$ is binary, taking values of -1 and 0, depending on whether a collision occurred or not; $S$ is the normalized deviation of the ego vehicle speed from the mean speed of the traffic; $D$ is related to the distance between the ego vehicle and the car in front, and takes the values of -1, 0 or 1 depending on whether the distance is \textit{close}, \textit{nominal} or \textit{far}, respectively. This term reflects the driver preference of having as much headway as possible. $E$ is the effort variable taking the value of 0 if the action is maintain; a value of -0.25, if the action is accelerate or decelerate; and -0.5 if it's hard accelerate or hard decelerate. $E$ receives -1 if the driver changes lane.

\begin{remark}
\textit{The terms of the reward function provided in \eqref{eq:rewardTr} are determined based on the specific problems that need to be solved to obtain driver dynamics. The solution of these problems reflects the driver\textquotesingle s preferences. Specific issues to be solved can be listed as collision avoidance (term $C$), performance (term $S$), safety (term $D$) and comfort or energy conservation (term $E$). Instead of discrete valued terms used in \eqref{eq:rewardTr}, continuously varying terms could be employed. However, while this would reflect the driver preferences in a finer manner, it could introduce unnecessary complexity. A more systematic selection of the reward function can be conducted for problems such as imitation learning, where inverse reinforcement learning methods are applied \citep{sutton2018reinforcement}}. 
\end{remark}

\subsection{Physical models}
For the modeled traffic scenario, simple kinematic models are used to obtain vehicle motion. Once the driver command is received, velocity and position dynamics are obtained as
\begin{eqnarray}
	x(t+\Delta t) &=& x(t) + v_{x}(t)*\Delta t+\frac{1}{2} a(t)\Delta t^{2} \nonumber \\
	y(t+\Delta t) &=& y(t) + v_{y}(t)*\Delta t \nonumber \\  
	v_{x}(t+\Delta t) &=& v_{x}(t) + a(t)*\Delta t,
\end{eqnarray}
where $v_x$ and $v_y$ are the velocity components in the $x$ and $y$ directions, $x$ and $y$ coordinates are the same as given in Figure~\ref{fig_model}, and $\Delta t$ is the simulation time step. It is noted that the dynamics are not discrete but continuous, and have to be approximated due to computer implementation.

\subsection{Training, average rewards and entropy} 
\begin{figure}[htb]
		\centering
		\begin{subfigure}[b]{0.4\textwidth}
			\includegraphics[width=\linewidth]{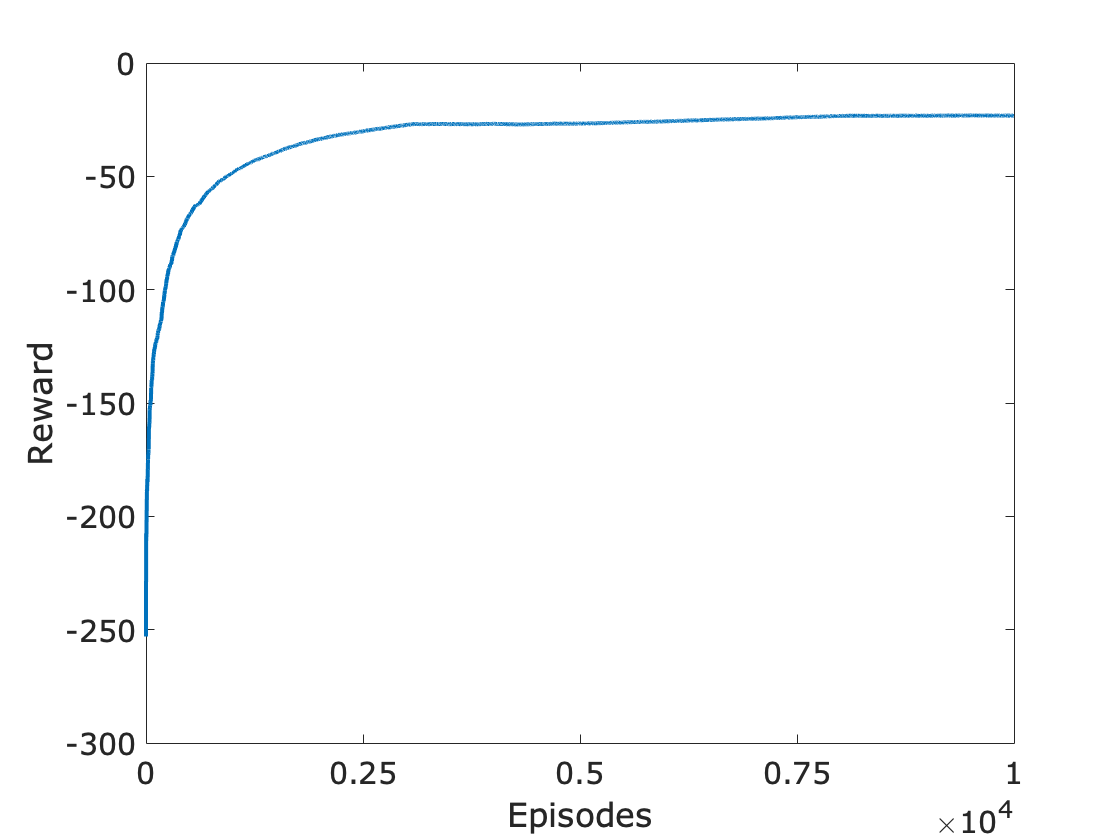}
			\caption{Average reward per episode in level-1 training}
			\label{fig_reward1}
		\end{subfigure} 
		~
		\begin{subfigure}[b]{0.4\textwidth}
			\includegraphics[width=\linewidth]{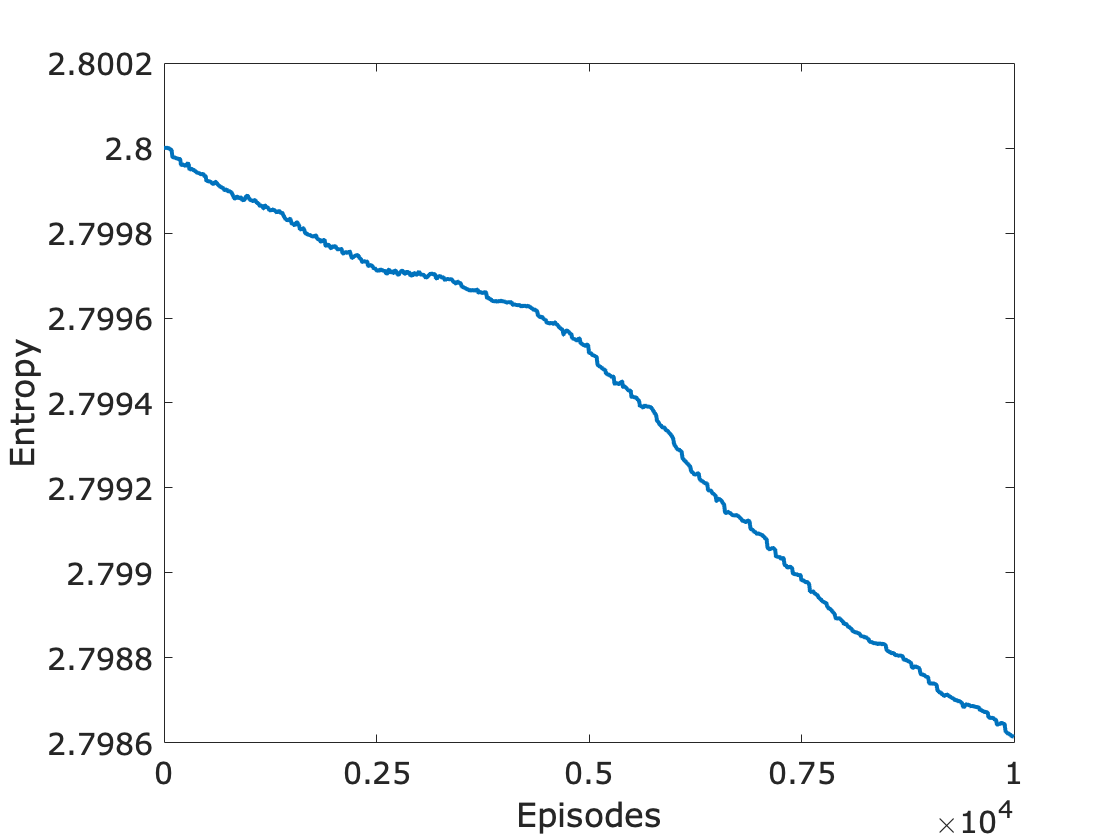}
			\caption{Entropy per episode in level-1 training}
			\label{fig_entropy1}
		\end{subfigure}
		\caption{Level-1 training}\label{fig:lev1}
	\end{figure}

\begin{figure}[htb]
		\centering
		\begin{subfigure}[b]{0.4\textwidth}
			\includegraphics[width=\linewidth]{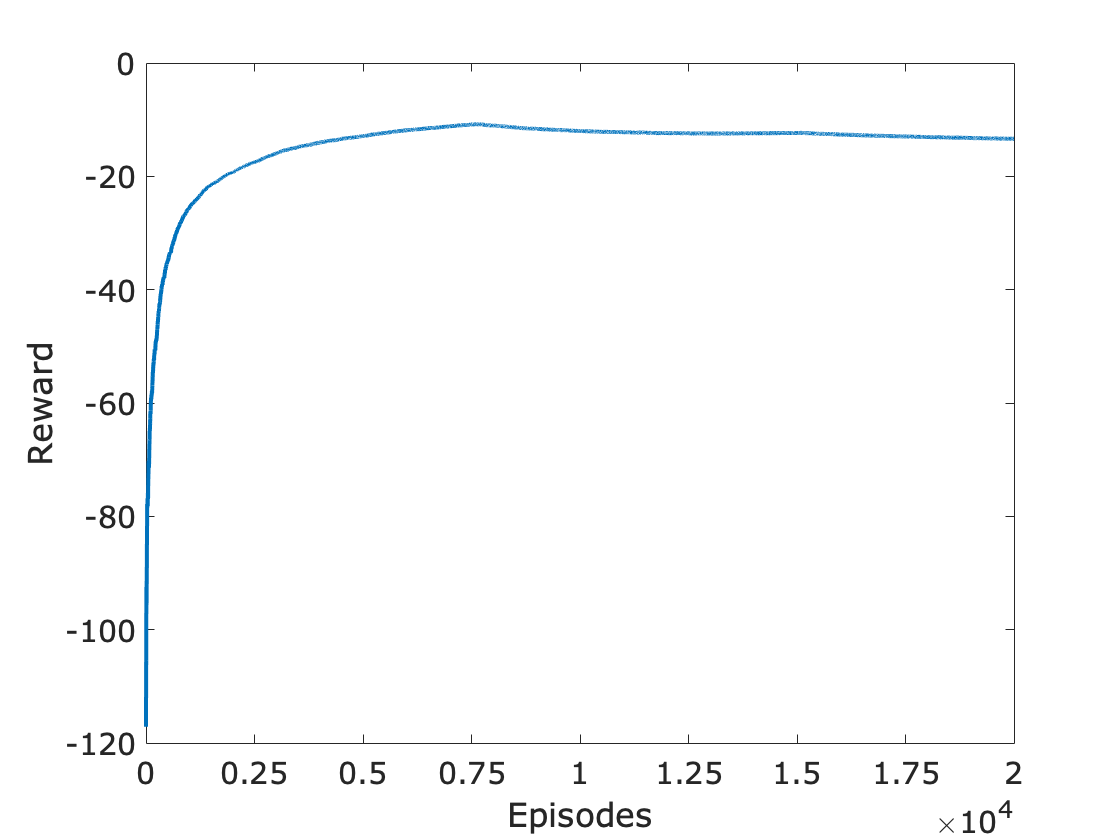}
			\caption{Average reward per episode in level-2 training}
			\label{fig_reward2}
		\end{subfigure} 
		~
		\begin{subfigure}[b]{0.4\textwidth}
			\includegraphics[width=\linewidth]{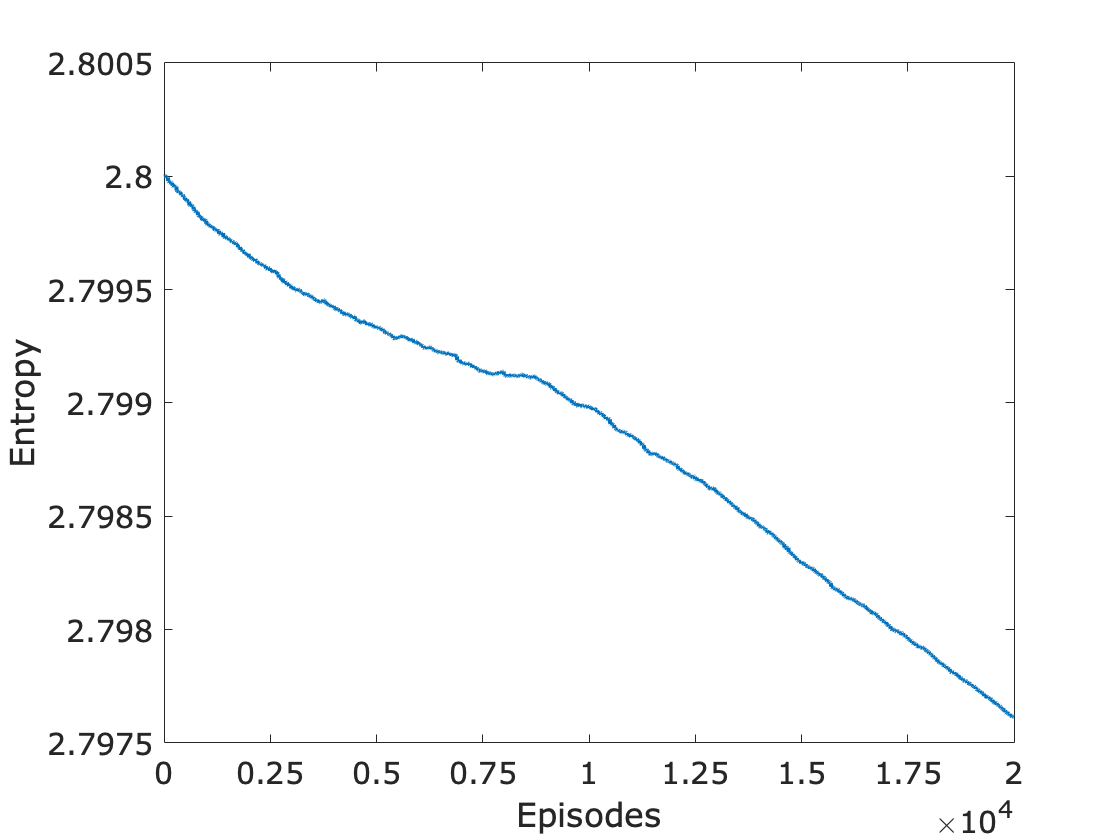}
			\caption{Entropy per episode in level-2 training}
			\label{fig_entropy2}
		\end{subfigure}
		\caption{Level-2 training}\label{fig:lev2}
	\end{figure}

\begin{figure}[H]
		\centering
		\begin{subfigure}[b]{0.4\textwidth}
			\includegraphics[width=\linewidth]{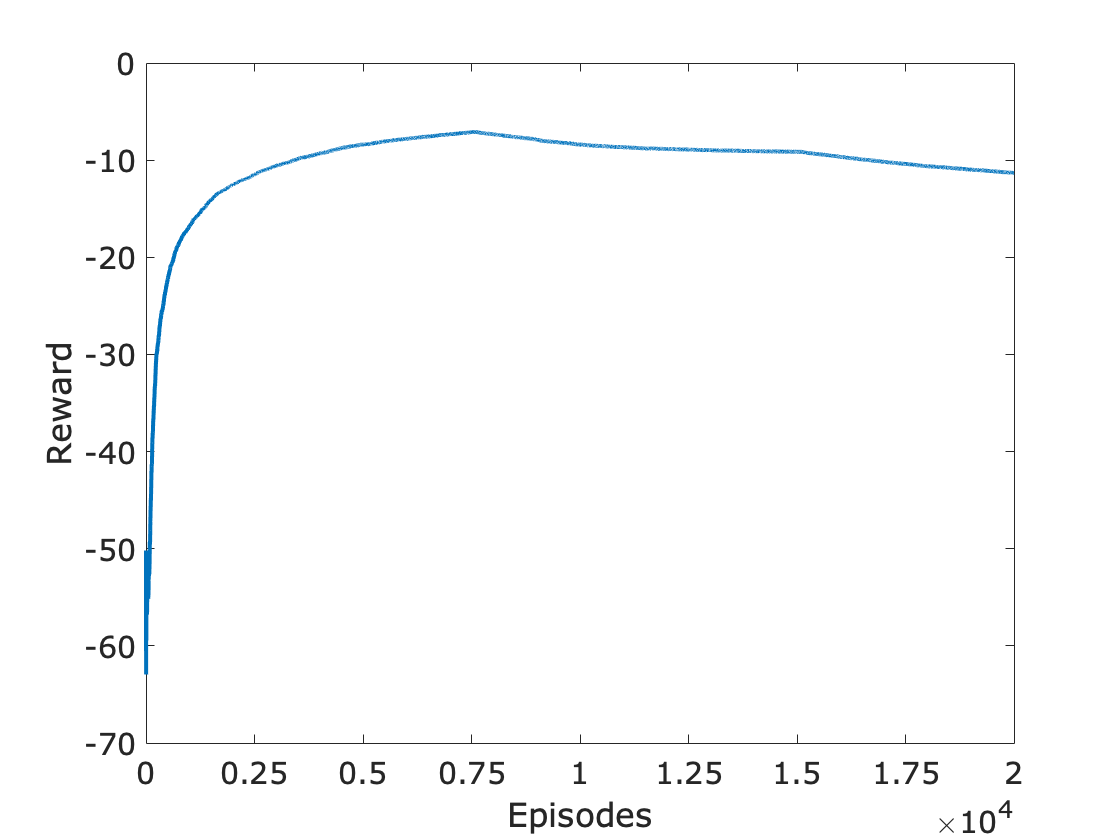}
			\caption{Average reward per episode in level-3 training}
			\label{fig_reward3}
		\end{subfigure} 
		~
		\begin{subfigure}[b]{0.4\textwidth}
			\includegraphics[width=\linewidth]{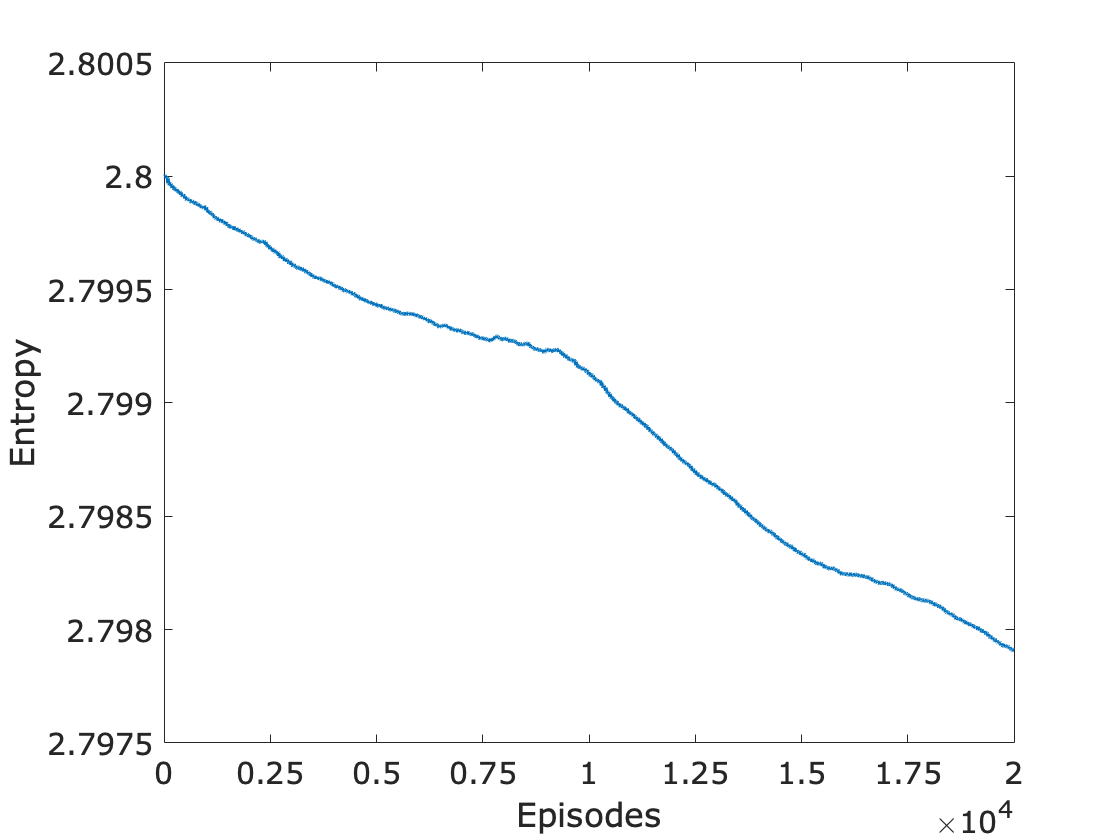}
			\caption{Entropy per episode in level-3 training}
			\label{fig_entropy3}
		\end{subfigure}
		\caption{Level-3 training}\label{fig:lev3}
	\end{figure}

\begin{figure}[htb]
		\centering
		\begin{subfigure}[b]{0.4\textwidth}
			\includegraphics[width=\linewidth]{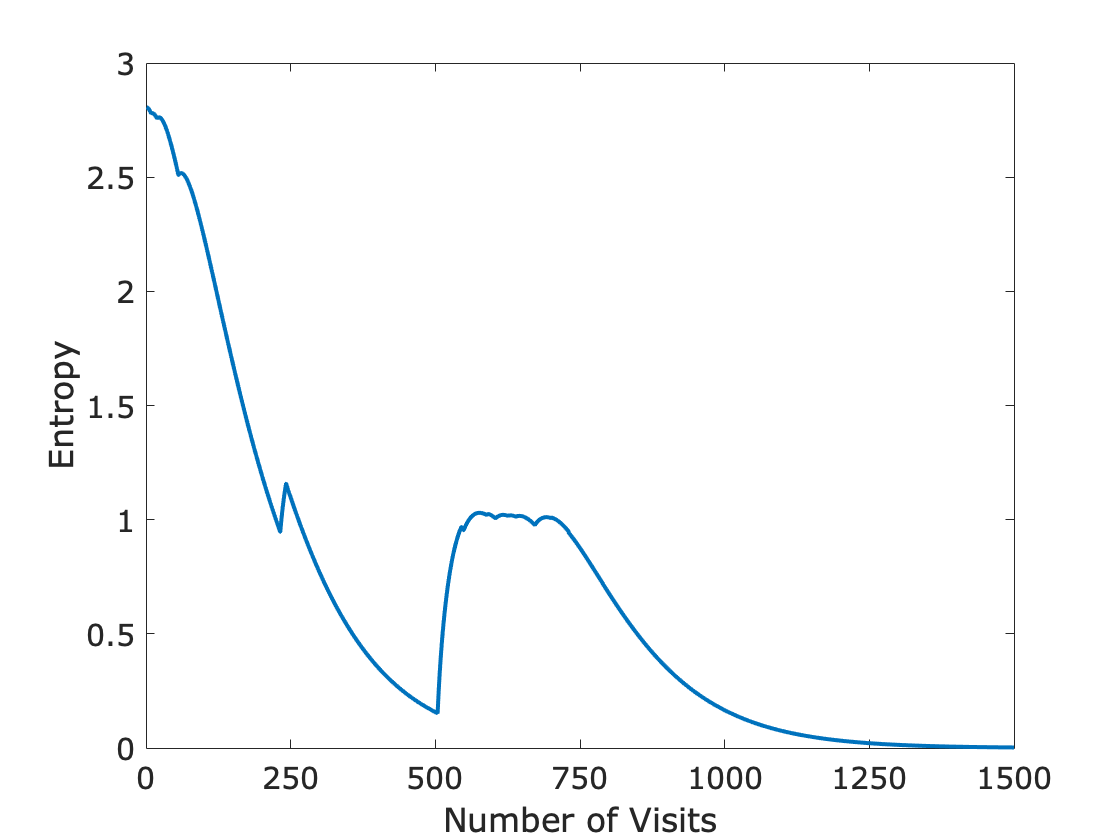}
			\caption{Entropy per visit of a state}
			\label{fig_entropystate1}
		\end{subfigure}
		~ 
		\begin{subfigure}[b]{0.4\textwidth}
			\includegraphics[width=\linewidth]{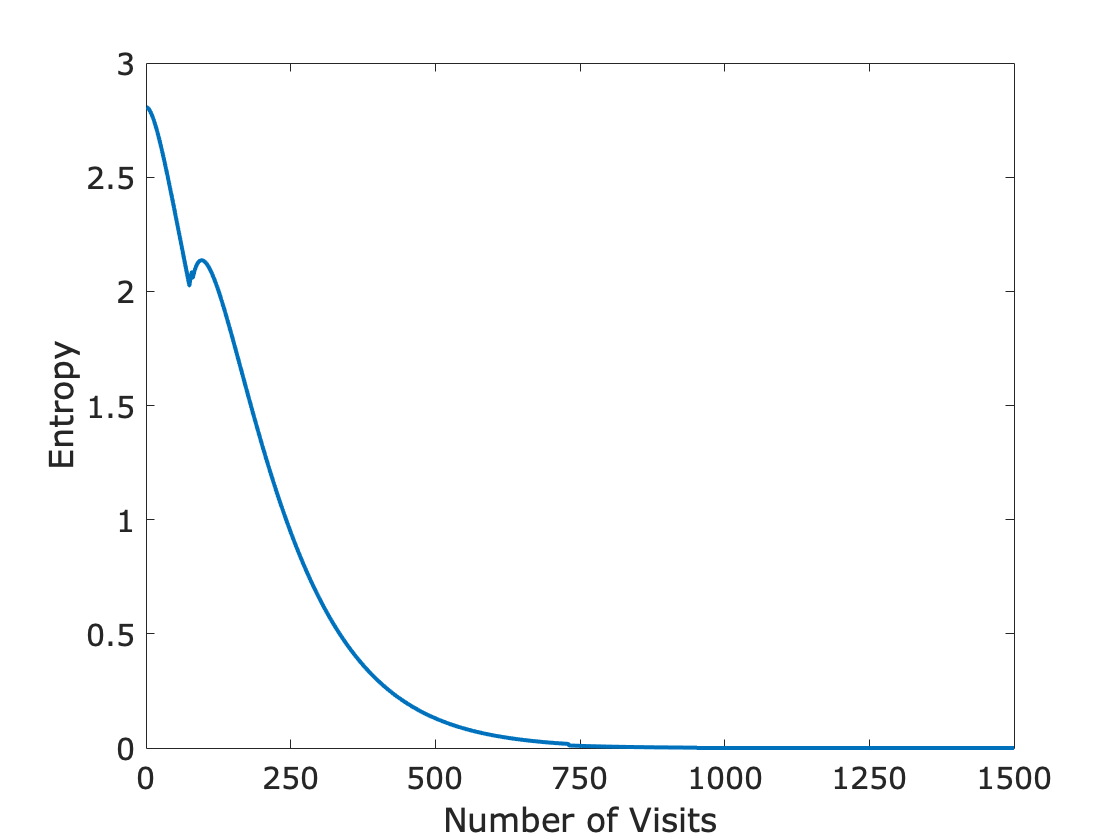}
			\caption{Entropy per visit of a different state}
			\label{fig_entropystate2}
		\end{subfigure}
		\caption{Entropy per visit plots of two randomly selected states.}\label{f:ent}
	\end{figure}

Driver policies with various levels are trained using the Jaakkola reinforcement learning algorithm explained in Section~\ref{sec:Jaak}. During training, up to 125 vehicles are used in the 5-lane traffic scenario. Figures \ref{fig:lev1}-\ref{fig:lev3} show the time evolution of average rewards during the training of level-1, level-2 and level-3 policies, together with the average entropy of the probability distribution over actions. The entropy of a probability distribution is calculated as
\begin{equation}
E= -\sum_{i=1}^n p_i \log(p_i),
\end{equation}
where $n$ is the number of probability values $p_i$. Since the drivers have 7 actions, we have $n=7$. The entropy is highest when the distribution is uniform, which is the case in the beginning of training, and drops as the training progresses and the distribution gets away from uniform. However, as seen from the figures, although the average reward converges relatively fast, entropy continues to drop at a very slow rate. The reason for this is that there are several states with lower probability of being visited during training and after a certain driving pattern emerges, the effect of these rarely visited states becomes very small. This can also be observed from Figure~\ref{f:ent}, where the entropies of the two frequently visited states are shown to converge to very small values. 

\subsection{Data validation}
The driver policies obtained using the game theoretical framework are compared with real traffic data provided by \cite{Colyar:07}. The data provides acceleration values of the drivers at each measurement instant. This data is first processed to obtain the states the drivers are in and then to find the frequency of taken actions at each state. These state-action frequency distributions form the real driver policies. To compare the distributions modeled using the proposed game theoretical approach and the distributions obtained from data, we use the Kolmogorov-Smirnov (KS) Test for Discontinuous Distributions \citep{conover1972kolmogorov}. In this test, if the unknown discrete probability distribution function is $F(x)$ and the hypothesized distribution is $H(x)$, the null hypothesis $H_0$ is defined as
\begin{equation}
H_0: F(x) = H(x)~for~all~x.
\end{equation}
Three test statistics used in the test are 
\begin{equation} \label{eq:teststat1}
D = sup_x |H_c(x) - S_n(x)|,
\end{equation}
\begin{equation} \label{eq:teststat2}
D ^-= sup_x (H_c(x) - S_n(x))
\end{equation}
and
\begin{equation} \label{eq:teststat3}
D^+ = sup_x (S_n(x) - H_c(x)),
\end{equation}
where $S_n(x)$ and $H_c(x)$ are the cumulative distribution functions (CDF) of the observed data and of the hypothesized distribution (model), respectively. The observed values of the test statistics $D$, $D^-$ and $D^+$ are defined as $d$, $d^-$ and $d^+$, respectively. The goal of the KS test is to calculate the probability of observing at least the value $d$ for the test statistics $D$, which can be stated as $P(D \geq d)$, given that the null hypothesis is true. This is achieved by first calculating $P(D^+ \geq d)$ and $P(D^- \geq d)$, and then obtaining  $P(D \geq d)$ as
\begin{equation} \label{eq:testprob}
P(D \geq d) = P(D^+ \geq d^+) + P(D^- \geq d^-).
\end{equation}
The details of obtaining $P(D^+ \geq d)$ and $P(D \geq d^-)$ are omitted here for brevity and can be found in \citep{conover1972kolmogorov}. Once $P(D \geq d)$ is calculated using \eqref{eq:testprob}, the null hypothesis is rejected if $P(D \geq d) \leq 0.05$. Since this test provides meaningful results for distributions with non-zero entries, action probabilities that are lower than 0.01 are set to 0.01 with normalization, for both the real data and the driver model.

Data validation is conducted individually for each driver: For the driver of interest, first, the action probability distributions for each visited state are computed. Second, these distributions are compared with derived driver policies using the proposed modeling framework via KS test. Specifically, the distributions from the data are compared with level-1, level-2 and level-3 policies. Finally, the percentage of the states that can be successfully modeled are reported. This procedure is repeated for every driver whose traffic data is available at \cite{Colyar:07}. The states that are visited less than a certain threshold number, $n_{limit}$, during actual driving by the drivers and during training of the policies are not taken into account. The results are reported for different values of $n_{limit}$. The algorithm used for this validation procedure is provided in Algorithm \ref{alg:KS}, where $n_{state}$ is the number of visited states by the driver, $n_{Vdriver}$ is the number of times the driver of interest visited the state being evaluated, $n_{Vmodel}$ is the number of times the state is visited during the training of the driver model, $n_{comp}$ is the number of states that are used in the comparison and $n_{success}$ is the number of states for which the null hypothesis is not rejected.
\begin{algorithm}[H] 
	\caption{Procedure of Comparison between one driver-one policy - Kolmogorov Smirnov}
	\begin{algorithmic}[1]
		\FOR {i = 1 to $n_{state}$} 
			\IF{$n_{Vdriver}  \geq n_{limit}$ and $n_{Vmodel}  \geq n_{limit}$ }
			
				\STATE $n_{comp}+=1$
				\STATE Set $p_i$ to the probability mass function given by the driver model for the state being evaluated.
				\STATE Set $k_i$ to the probability mass function calculated from the driver data for the state being evaluated.
					\STATE Set $H_c$ to the cumulative  distribution function obtained from $p_i$. 
					\STATE Set $S_n$ to the cumulative  distribution function obtained from $k_i$.
					\STATE Test the null hypothesis using KS test.
				
				\IF{Null hypothesis is not rejected}
					\STATE $n_{success} += 1$
				\ENDIF
			\ENDIF
		\ENDFOR
		
		\STATE Set the percentage of the successfully modeled states to  $n_{success} /n_{comp}$.
		
	\end{algorithmic}
	\label{alg:KS}
\end{algorithm}

Data validations using Algorithm \ref{alg:KS} are conducted for $n_{limit}$ values of 0, 3 and 5, for each individual driver and the results are reported below.

\subsubsection{Validation with $n_{limit}=1$}

\begin{figure}[htb]
	\centering
	\includegraphics[width=4.1in]{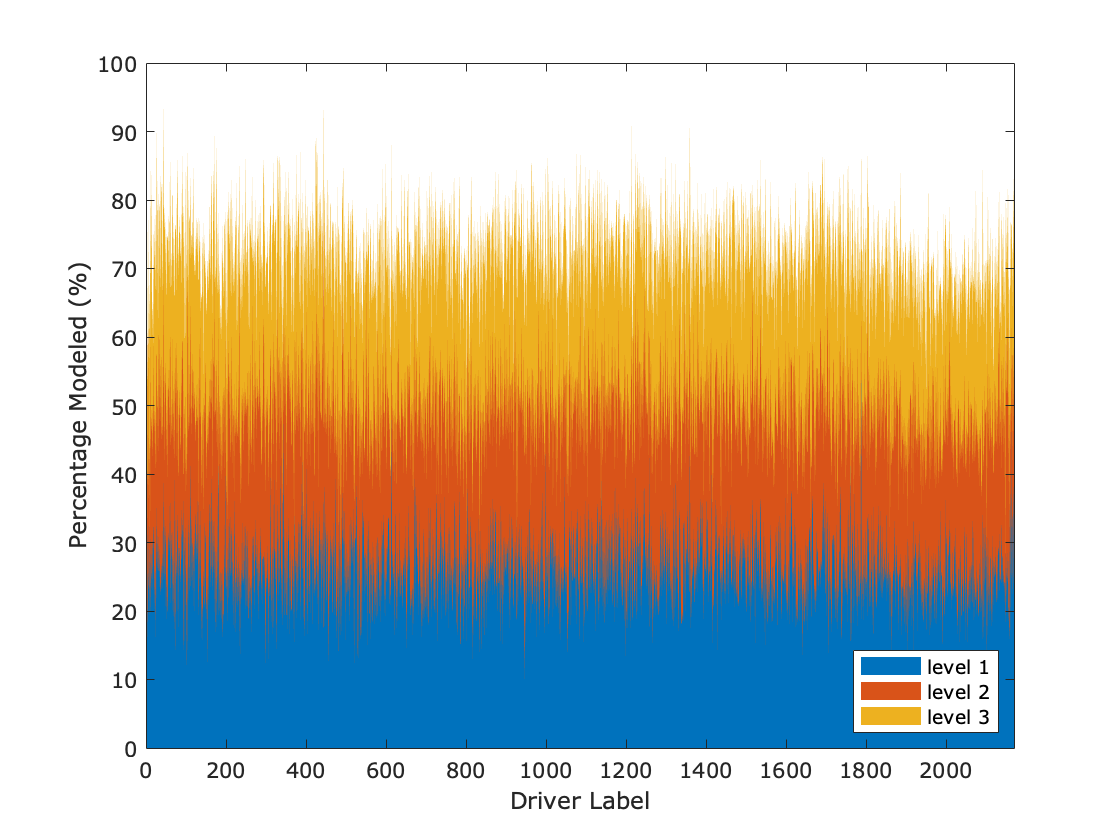}
	\caption{Percentages of successfully modeled states for each driver, using level-k models, when $n_{limit}=1$.}
	\label{kspd1}
\end{figure}

\begin{figure}[H]
	\centering
	\includegraphics[width=4.1in]{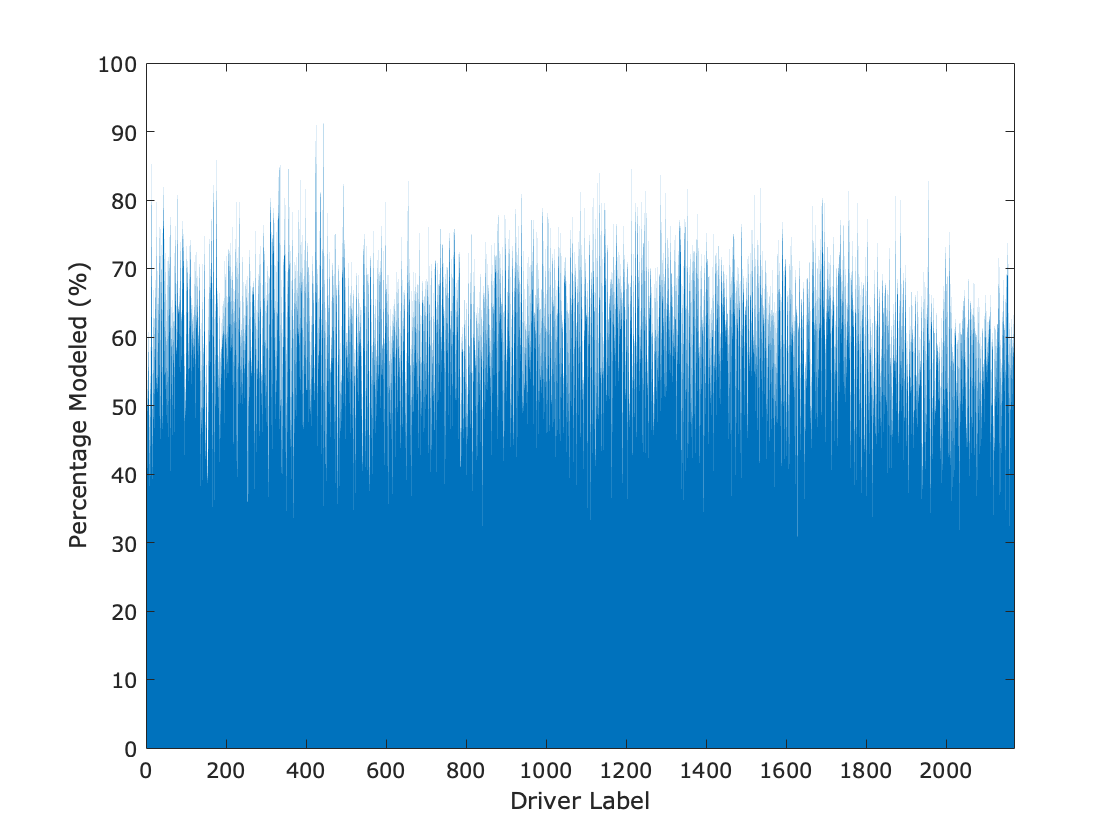}
	\caption{Percentages of successfully modeled states for each driver, using the uniform distribution model, when $n_{limit}=1$}
	\label{kspd2}
\end{figure}

\begin{figure}[htb]
	\centering
	\includegraphics[width=4.1in]{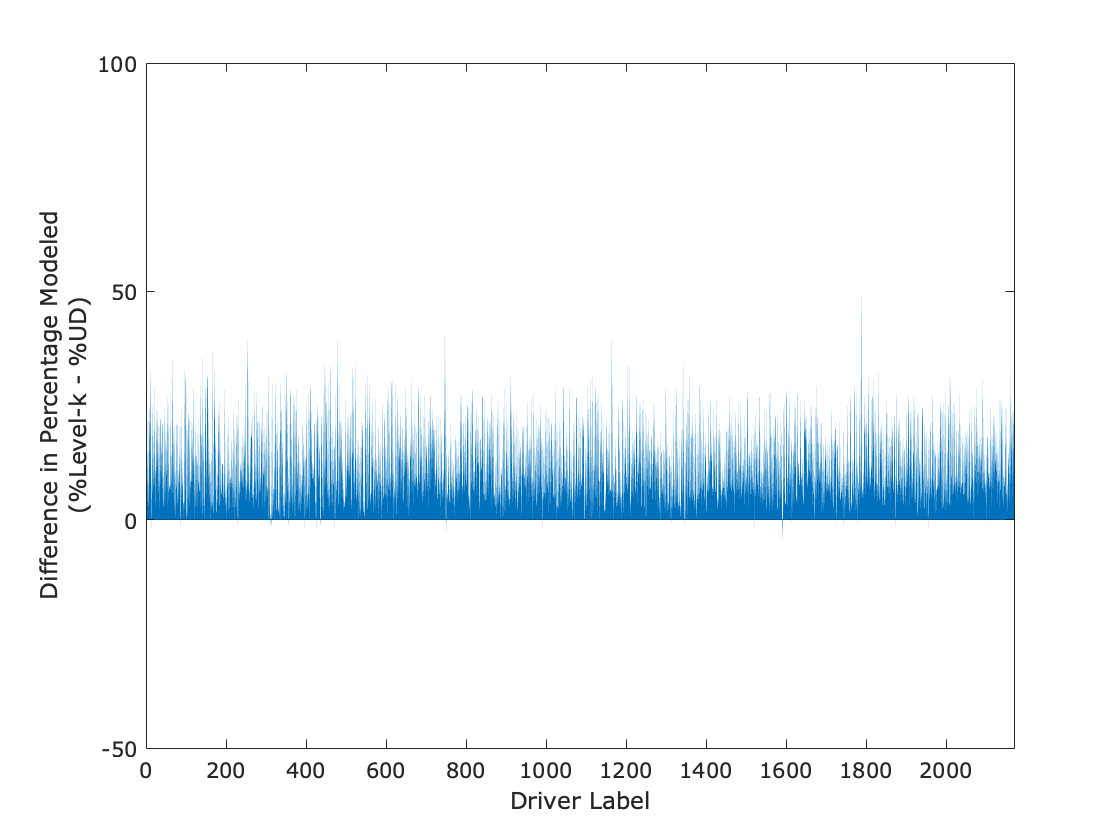}
	\caption{Difference in successfully modeled state percentages, for each driver, using level-k (combined) and UD models, when $n_{limit}=1$.}
	\label{kspd3}
\end{figure}

Figure \ref{kspd1} shows the percentage of successfully modeled states for each driver, using level-1, level-2 and level-3 policies as the driver models. Each vertical thin line represents an individual driver and the x-axis shows the driver labels. It is seen that there are more than two thousand drivers in the real traffic data. The y-axis shows the percentage of successfully modeled states. The colors blue, red and yellow represent the successfully modeled states by level-1, level-2 and level-3 policies, respectively. Figure \ref{kspd2} also shows the successfully modeled states, but this time using a uniform distribution (UD) model for the drivers. Finally, Figure~\ref{kspd3} shows the difference between the level-k models (combined) and the UD model, in terms of modeling percentages. It is seen that the level-k models cumulatively perform better than the UD model. It is noted that although the level-k models perform better, the percentage of states that are modeled using the UD model are high. The main reason for this is the small $n_{limit}$, which is 1 in this case. For the state that is visited only 1 time by the actual driver, the KS test does not reject the hypothesis that the action distribution over this state, obtained from the data, is sampled from a UD, since the sample size is not large enough to arrive at a conclusion.

\begin{figure}[htb]
	\centering
	\includegraphics[width=4.1in]{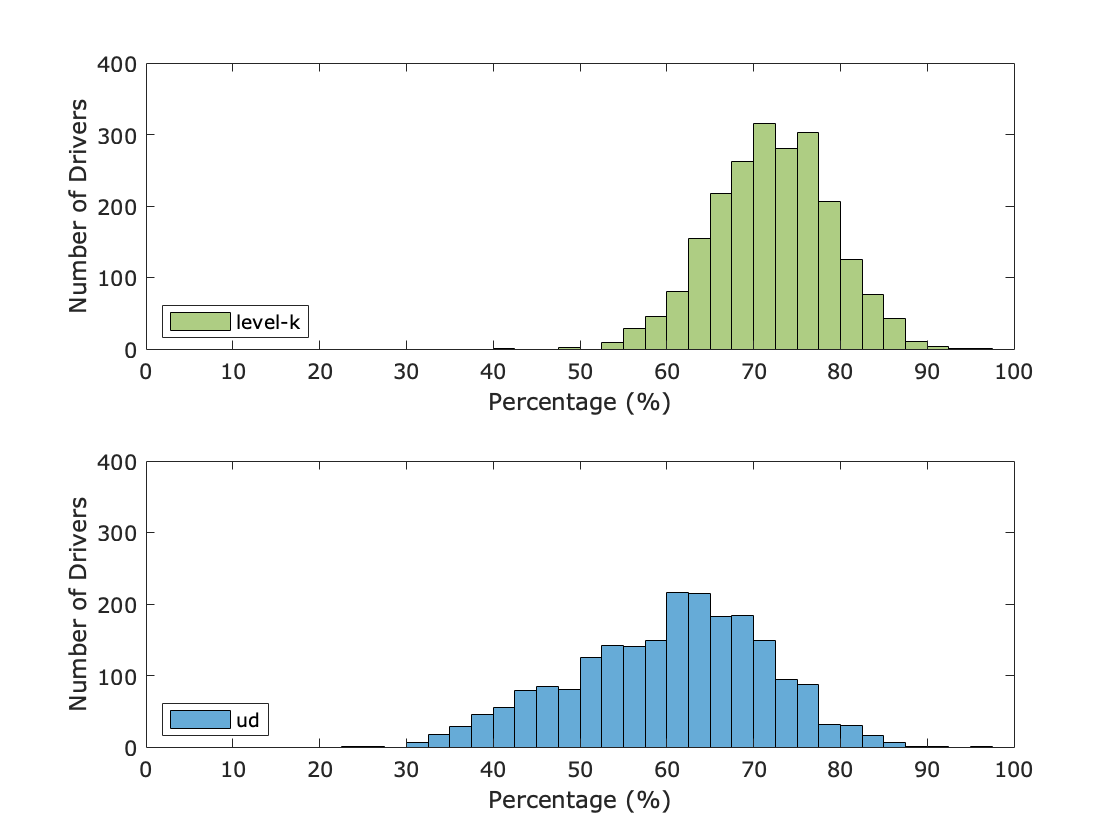}
	\caption{Distribution of modeled percentages for combined policies and dumb policy, when $n_{limit}=3$.}
	\label{kspd5}
\end{figure}

Figure \ref{kspd5} shows the number of drivers, on the y-axis, and the percentage of the successfully modeled states by the level-k models (top) and the UD model (bottom). For example, it shown that around 80\% of the total visited states of 300 drivers are successfully modeled by the level-k models, while only a handful of drivers' states are successfully modeled up to 60\% by the same models. The more the distribution of bars on these figures are grouped on the right, the better, since it shows that successfully modeled state percentages are higher.

\subsubsection{Validation with $n_{limit}=3$}

\begin{figure}[htb]
	\centering
	\includegraphics[width=4.1in]{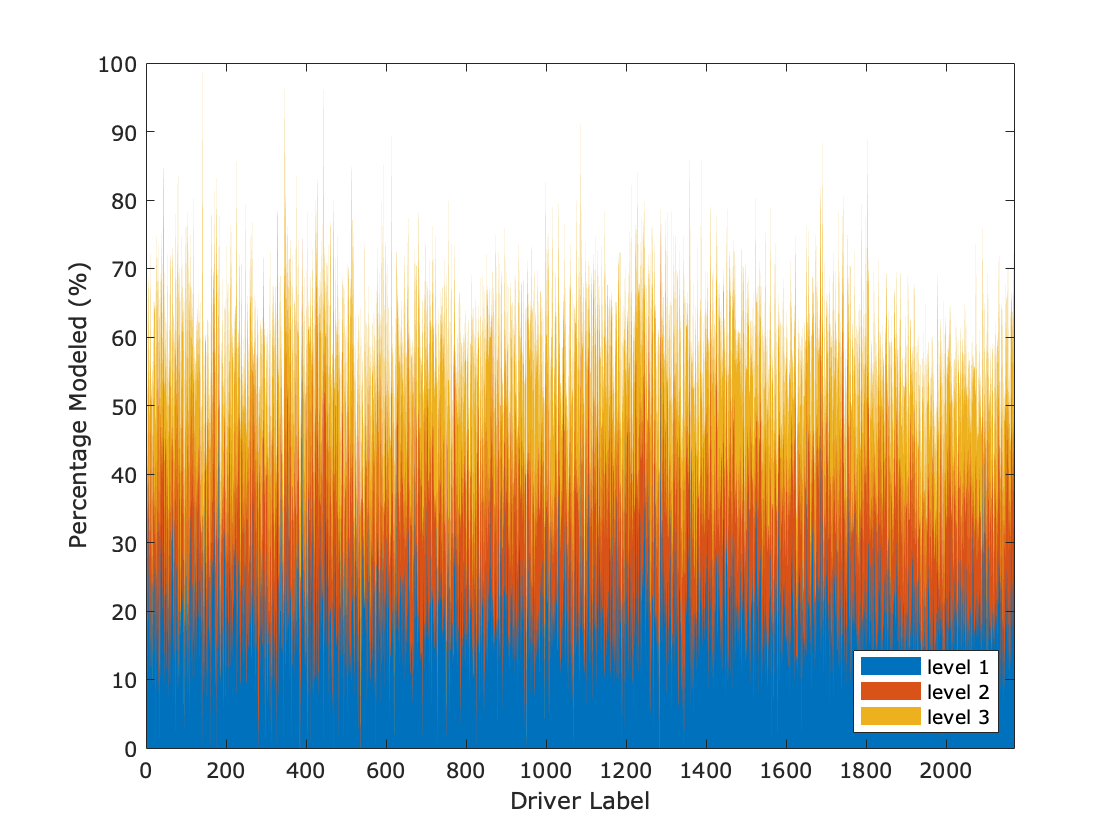}
	\caption{Percentages of successfully modeled states for each driver, using level-k models, when $n_{limit}=3$.}
	\label{ks3d1}
\end{figure}

\begin{figure}[H]
	\centering
	\includegraphics[width=4.1in]{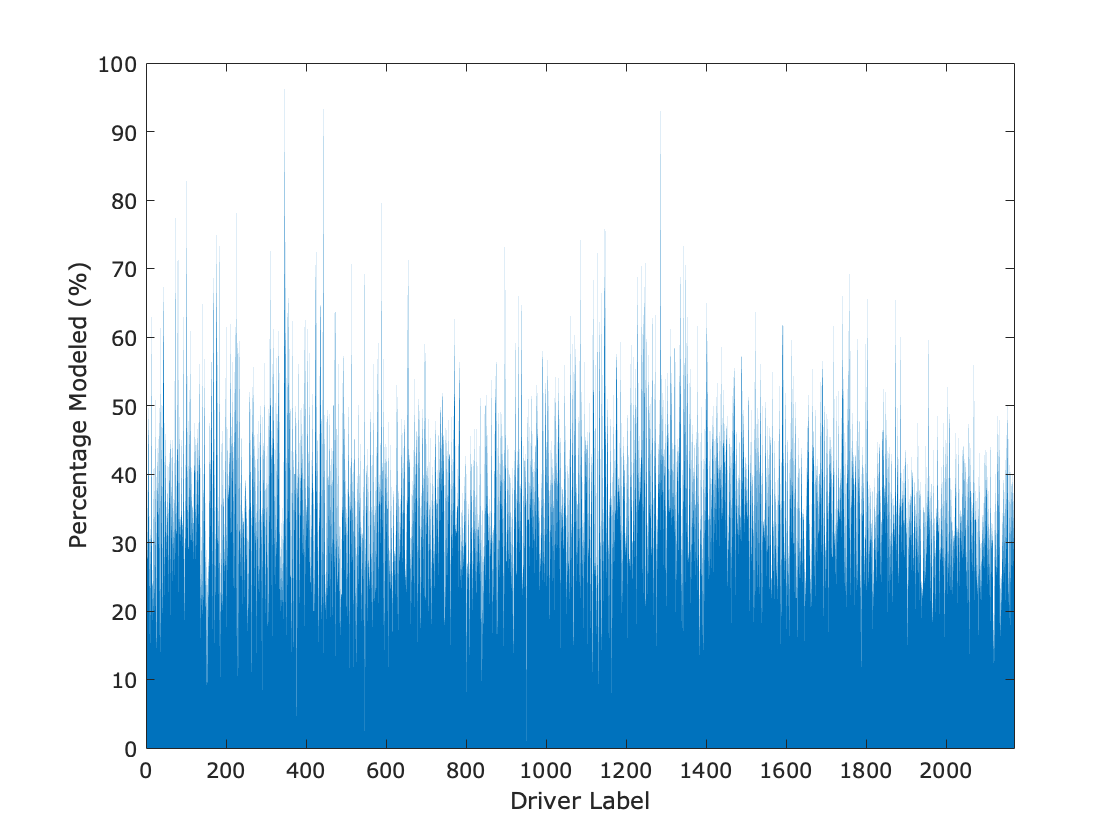}
	\caption{Percentages of successfully modeled states for each driver, using the uniform distribution model, when $n_{limit}=3$.}
	\label{ks3d2}
\end{figure}

\begin{figure}[H]
	\centering
	\includegraphics[width=4.1in]{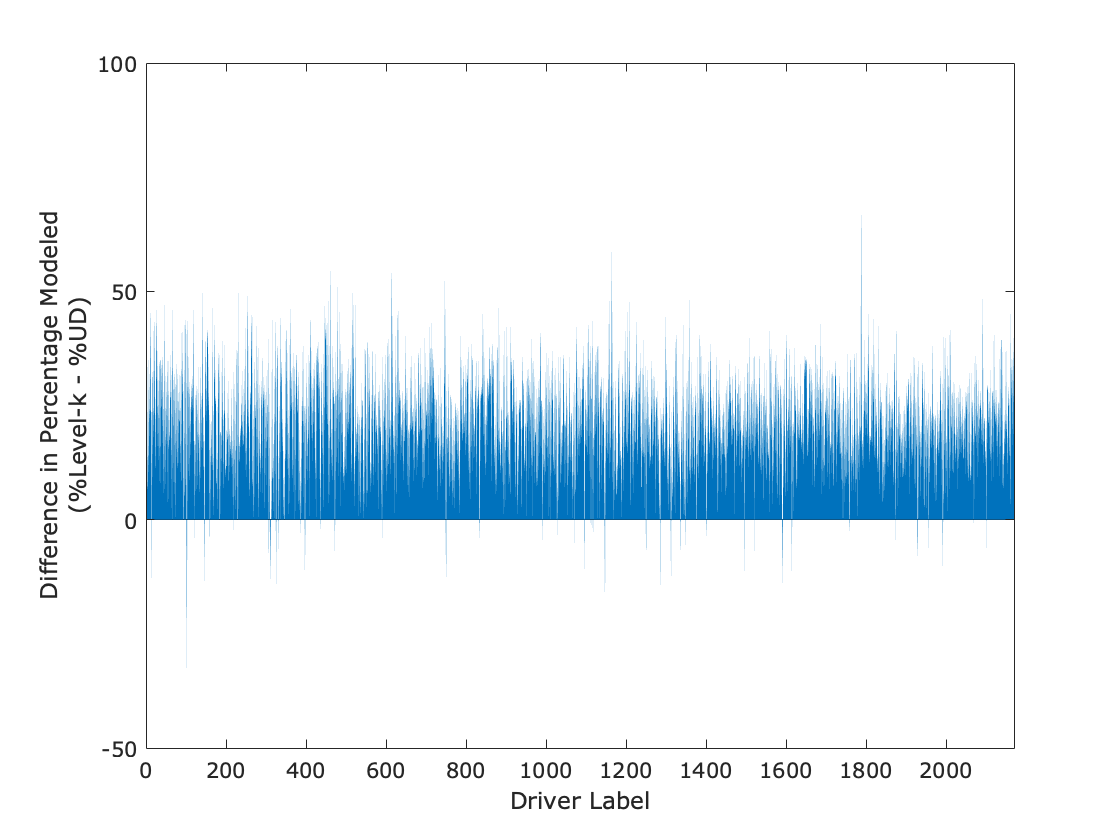}
	\caption{Difference in successfully modeled state percentages, for each driver, using level-k (combined) and UD models, when $n_{limit}=3$.}
	\label{ks3d3}
\end{figure}

Figure \ref{ks3d1} and \ref{ks3d2} show the percentages of the successfully modeled states by the level-k models and the UD model, respectively. As Figure~\ref{ks3d3} demonstrates, with an increased threshold ($n_{limit}=3$) the better performance of level-k models, compared to the UD model, becomes more prominent . This is due to the increased power of the KS test with the elimination of rarely visited states by the drivers.


\begin{figure}[htb]
	\centering
	\includegraphics[width=4.1in]{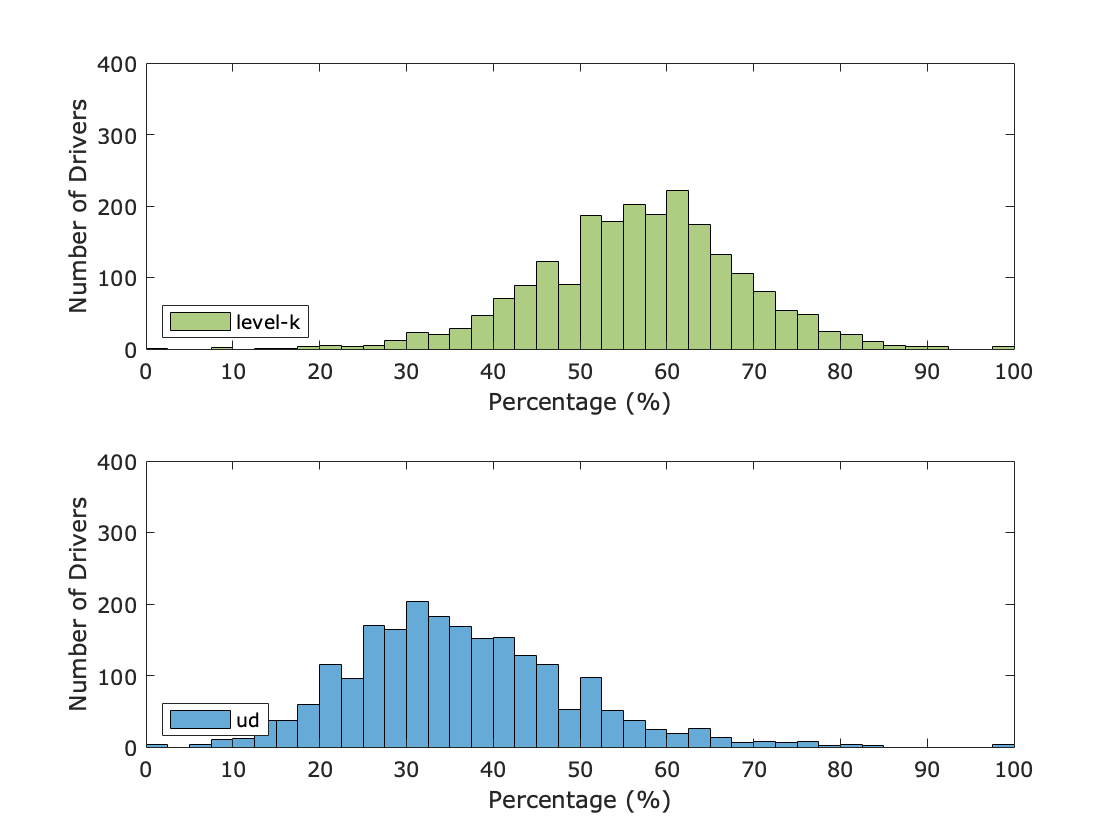}
	\caption{Distribution of modeled percentages for combined policies and dumb policy, when $n_{limit}=3$.}
	\label{ks3d5}
\end{figure}

Figure \ref{ks3d5} also demonstrates the increased performance difference between the level-k and the UD models, in favor of the level-k, by showing the distribution of the number of drivers over successfully modeled state percentages. As more of the rarely visited states
are removed from the comparison, the KS test is able to reject the null hypothesis more frequently for the UD models.

\subsubsection{Validation with $n_{limit}=5$}

\begin{figure}[htb]
	\centering
	\includegraphics[width=4.1in]{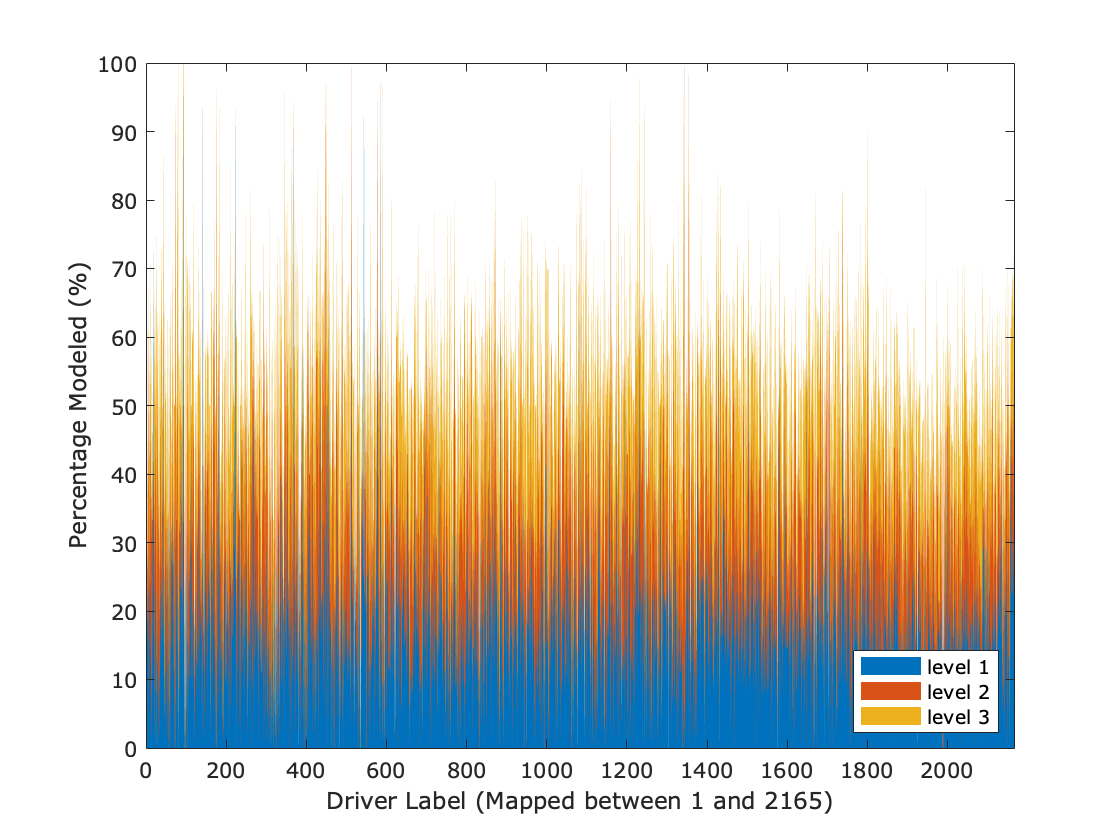}
	\caption{Percentages of successfully modeled states for each driver, using level-k models, when $n_{limit}=5$.}
	\label{ks5d1}
\end{figure}

\begin{figure}[H]
	\centering
	\includegraphics[width=4.1in]{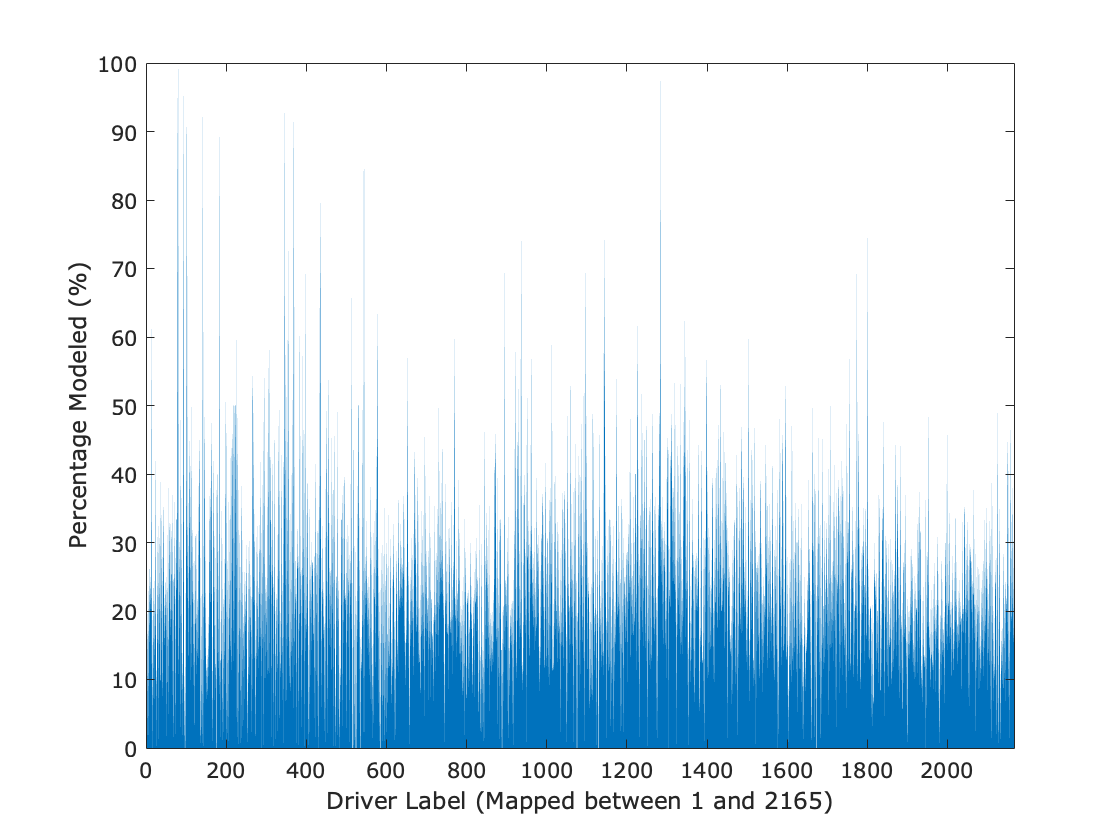}
	\caption{Percentages of successfully modeled states for each driver, using the uniform distribution model, when $n_{limit}=5$.}
	\label{ks5d2}
\end{figure}

\begin{figure}[htb]
	\centering
	\includegraphics[width=4.1in]{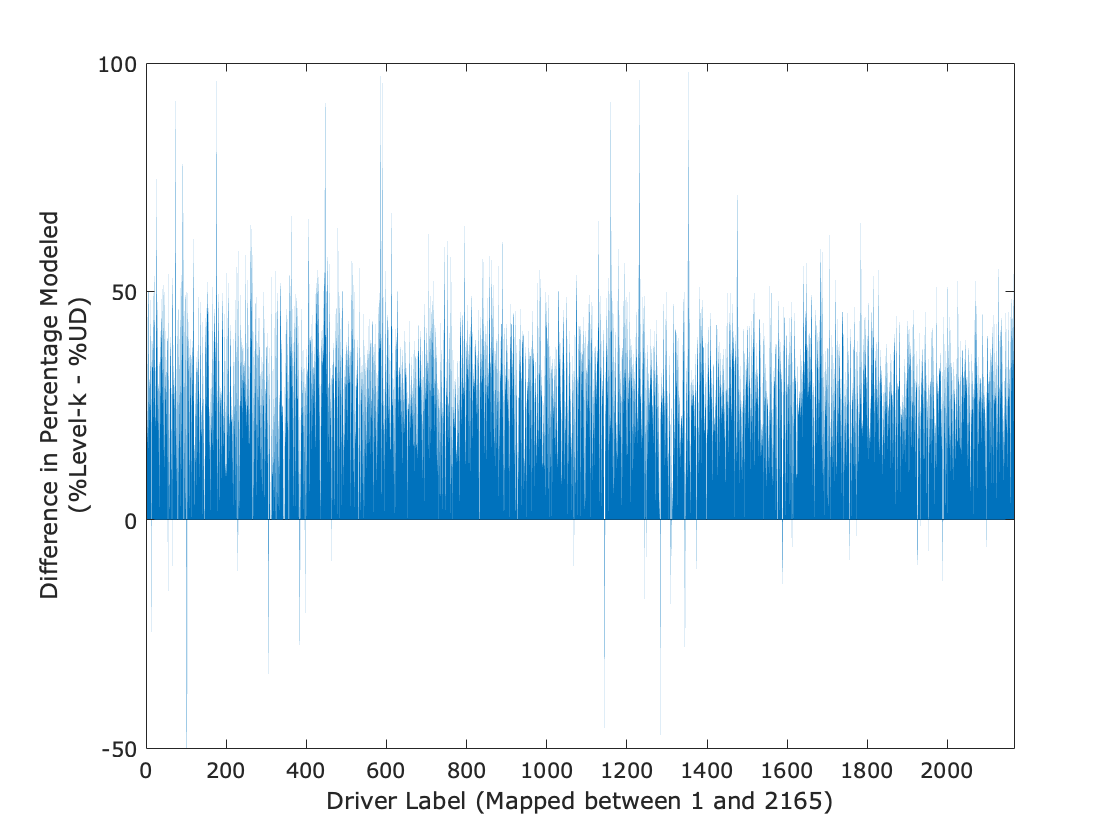}
	\caption{Difference in successfully modeled state percentages, for each driver, using level-k (combined) and UD models, when $n_{limit}=5$.}
	\label{ks5d3}
\end{figure}

\begin{figure}[H]
	\centering
	\includegraphics[width=4.1in]{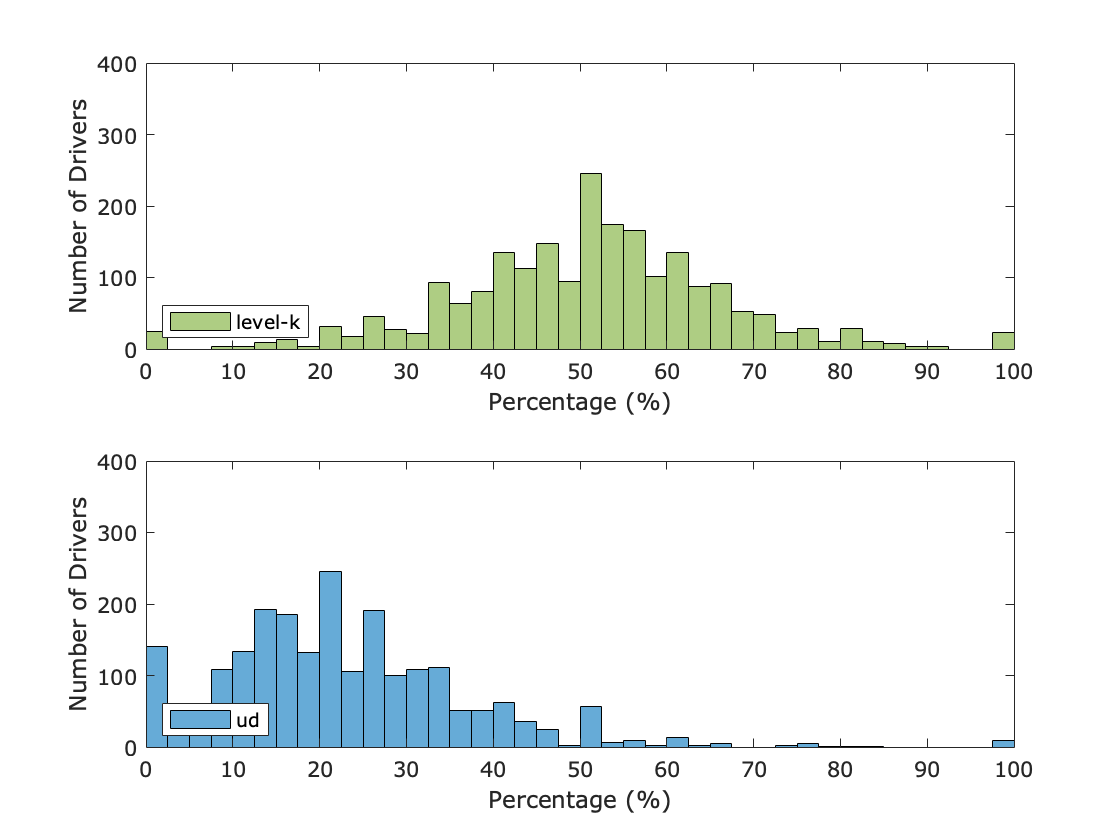}
	\caption{Distribution of modeled percentages for combined policies and dumb policy, when $n_{limit}=5$.}
	\label{ks5d5}
\end{figure}

When even more of the rarely visited states are removed from the test by increasing the threshold further to $n_{limit}=5$, the same trend in increased performance difference between the level-k models and the UD model continues, as seen in Figures \ref{ks5d1}, \ref{ks5d2},  \ref{ks5d3} and \ref{ks5d5}.



To summarize, KS test results demonstrate that with varying levels of success, the exploited game theoretical modeling framework provides driver models whose predictive power can be validated with real traffic data. Since it is hard to find similar driver models in the literature, with probability distributions over actions, the results are compared with a uniform distribution model. For all three $n_{limit}$ values, level-k policies (combined) performed better than the UD model.

\section{Computational Complexity} \label{sec:cc}
In both of the application areas, hybrid airspace modeling and road traffic modeling, discussed in the previous sections, the exploited game theoretic framework employed Jaakkola reinforcement learning method explained in Section \ref{sec:Jaak}. In this section, a computational cost analysis is provided for this method. Specifically, two questions are answered: 1) How do the action and observation space sizes affect the computational complexity? 2) How do the complexities of the physical motion models affect the computational complexity?

In the analysis, the dimensions of the state and action spaces are taken as $S$ and $A$, respectively. It is assumed that a) every state is visited $K$ times during one sweep, and b) the value functions are updated after each visit. For each state-action pair, $(s_i, a_i)$, the functions given below are calculated:

\begin{eqnarray}
\beta_t(s_i,a_i) &=& \left(1-\frac{\chi_t(s_i,a_i)}{K_t(s_i,a_i)}\right) \gamma_t\beta_{t-1}(s_i,a_i) + \frac{\chi_t(s_i,a_i)}{K_t(s_i,a_i)} \nonumber \\
\beta_t(s_i) &=& \left(1-\frac{\chi_t(s_i)}{K_t(a_i)}\right) \gamma_t\beta_{t-1}(s_i) + \frac{\chi_t(s_i)}{K_t(s_i)} \nonumber \\
Q_t(s_i,a_i) &=& \left(1-\frac{\chi_t(s_i,a_i)}{K_t(s_i,a_i)}\right)Q_{t-1}(s_i,a_i) + \beta_t(s_i,a_i)(R_t - R) \nonumber \\
V_t(s_i) &=& \left(1-\frac{\chi_t(s_i)}{K_t(s_i)}\right)V_{t-1}(s_i) + \beta_t(s_i)(R_t - R) \nonumber \\
for~each~action~a_k \neq a_i \nonumber \\
\beta_t(s_i,a_k) &=& \gamma_t\beta_{t-1}(s_i,a_k) \nonumber \\
Q_t(s_i,a_k) &=& Q_{t-1}(s_i,a_k) + \beta_t(s_i,a_k)(R_t - R) \nonumber \\
for~each~state~s_j \neq s_i ~and~action~a_l \nonumber \\
\beta_t(s_j,a_l) &=& \gamma_t\beta_{t-1}(s_j,a_l) \nonumber \\
\beta_t(s_j) &=&\gamma_t\beta_{t-1}(s_j) \nonumber \\
Q_t(s_j,a_l) &=& Q_{t-1}(s_j,a_l) + \beta_t(s_j,a_l)(R_t - R) \nonumber \\
V_t(s_j) &=& V_{t-1}(s_j) + \beta_t(s_j)(R_t - R) \nonumber \\
for~each~state~s_x~and~action~a_y \nonumber \\
\pi(a_y|s_x) &=& (1-\epsilon)\pi(a_y|s_x)+\epsilon\pi^1(a_y|s_x).
\end{eqnarray}
Therefore, for each state visit, $((24+(A-1)*(4))+(S-1)*(A)*(8)+S*A*4)$ operations are required. Visiting all of the states, $K$ times each, the total number of operations become, $K*S*((24+(A-1)*(4))+(S-1)*(A)*(8)+S*A*4)$, which can be expressed more compactly as
\begin{equation}
R = c_aS^2A + c_bSA + c_cS,
\end{equation}
where $R$ is the number of required operations, and $c_a,c_b$ and $c_c$ are constants. Therefore, the number of total operations can be given as $\mathcal{O}(|A||S|^2)$.

Using more complex vehicle dynamics, for example increasing the number of differential equations of the vehicle dynamics by $c$, will result in a computational cost that is $c$ times of the initial cost. Since this a constant effect, the total number of operations can still be expressed by $\mathcal{O}(|A||S|^2)$.

\section{Ongoing and Future work}
\label{sec:future}

In this section, we explain the current and future work about the game theoretical modeling method explained in this paper. Part of these studies are already formulated but not included in this article for a concise and clear exposition. Below, we briefly mention a few that have the potential to improve the existing work in a meaningful manner.

\subsection{3D hybrid airspace}
\label{s:3dhybrid}
Unmanned aircraft systems (UAS) integration into National Airspace System (NAS) studies presented in this article use a 2-dimensional (2D) geometry for aircraft motion. Although it is demonstrated that the proposed framework can be used to provide significant qualitative analysis power in UAS integration scenarios, the study is still limited and need to be extended to 3D airspace. Preliminary studies are conducted in this direction, details of which can be found in \cite{musavi20183d}. One important distinction in this study is the need for an approximate reinforcement learning algorithm to handle the dramatically increased state space due to the 3D geometry.

In the 2D case, the observation space consists of 9 variables. 6 of these variables can take 5 different values, while the remaining 3 variables can take 3 different values. Therefore, the size of the observation space is  $5^6 \times 3^3 = 421875$. This means that 421875 rows are required in the Q-table to represent each state. To store the values required during training, 16 columns are required. These columns are: state id, 
state value - V, state visit count, state beta, action 1 probability, action 1 Q value, action 1 count, action 1 beta, action 2 probability, action 2 Q value, action 2 count, action 2 beta, action 3 probability, action 3 Q value, action 3 count and action 3 beta. Hence, $421875 \times 16 
= 6750000$ values are stored in the Q-table as double. Since each double 
requires 8 bytes,  $54$ MB memory is required to store the Q table. This is not a significant amount of memory, which can be handled by any modern computer without any problem. However, when the geometry changes from 2D to 3D, the memory requirement becomes infeasible, even if we keep the action space the same. Adding 6 more observation states with 5 possible values for each, the dimension of the observation space becomes $5^{12} \times 3^3 = 6591796875$, which translates into a requirement of 800GB of memory to store the Q table. This calculation shows the necessity to use an approximate reinforcement learning method that eliminates the need for storing a Q table. The Neural Fitted Q-learning method, explained in Section~\ref{sec:nfq} is currently being tested for this task.

\subsection{Large scale cyber-security scenarios}
One ongoing study is creating a model of a large scale cyber-attack scenario, where multiple attackers try to hack into a cyber-physical system and several defenders try to keep the system safe. Reliable predictions of attacker-defender dynamics is valuable since they help design systems resilient to cyber-attacks. A two-person model of a cyber-attack scenario of a smart grid system is already conducted by \cite{Backhaus:13}. For a larger scenario, similar to the 3D airspace case, problems of increased state space should be solved together with integrating fast optimization algorithms to optimize the system design. Furthermore, reward function design for multiple attackers and defenders is a challenge especially if coordination within the attackers or defenders is envisioned. 

\subsection{Data validation}
Although data validation studies are presented in earlier sections, they are still at their initial stages due to several reasons. First, a lot more data is required for a reliable validation. For example, in UAS integration studies, validation is conducted using a data-validated model of manned aircraft encounters. We hope to obtain UAS and manned aircraft encounter data in the near future as the technology advances. Furthermore, in the traffic scenario, we used US101 data for validation but more road data is expected to be collected to ensure that the proposed model has the capability to model a large variation of highway configurations. In addition, as we use more data, we plan to fit the model parameters to certain amount of data and then validate with independent traffic data. It is noted that parameter fitting to data is not straightforward in the proposed method since reinforcement learning is involved at various reasoning levels. Second, new statistical goodness of fit methods need to be implemented to validate the model's power of prediction. To the best of our knowledge, no goodness of fit methods are implemented for either the UAS integration scenarios or road traffic scenarios where multiple actions are involved. In our ongoing work we are using Chi-square goodness of fit test and Kolmogorov goodness of fit test to validate the method. 

\section{Open problems and research opportunities}
\label{sec:open}

As mentioned in Section \ref{s:3dhybrid}, one of the limitations of the level-k thinking solution concept is that an agent's assumption about other agents' levels does not change during the game, and one solution to this may be the dynamic level-k approach. However, this is only a partial solution since the level types still remain unchanged. For example, if Agent~A, after watching Agent~B for a few time steps, decides that his or her assumption that Agent~B is a level-0 player is wrong, then Agent~A needs to update his or her assumption to another level. However, the set of levels he or she can choose from are fixed: Agent A can modify his assumption and assume that Agent~B is a level-1 player (instead of level-0), therefore the best response to this is producing the actions of a level-2 player. Agent-A can only change his or her assumption to level-k, where k=1,2..., n, where all levels are already trained and determined. A better, but more computationally expensive approach would be the following: After watching Agent~B for a few time steps, Agent~A can update his or her assumption by using, for example, a Bayesian update and come up a with a policy that is not in the pre-trained set of levels. After this update, Agent~A runs a reinforcement learning algorithm online to determine the best response to Agent~B's newly updated policy. To achieve this, we need to find new software and hardware solutions to handle the computational demand. 

Another open problem from a control engineering point of view is the problem of stability. Stability in RL is already being studied by the control community as an open problem \citep{bucsoniu2018reinforcement}. When used in collaboration with game theory, specifically level-k thinking, the problem becomes even harder to solve, and thus presents a rewarding research direction.

\section{Summary}
\label{sec:summary}

In this article, we reviewed a modeling approach where reinforcement learning and game theory work in tandem to predict cyber-physical human system (CPHS) behavior. Starting from the basic building blocks, we explained the method in detail and then presented two cases where models are created for unmanned aircraft systems (UAS) integration into National Airspace (NAS) and highway traffic scenarios. In both cases, validation studies were discussed using different methods, including using real world data. Finally, we presented ongoing and future works, together with related open problems, which can serve as fruitful research directions.

\section*{Acknowledgements}
\label{sec:ack}
This effort was sponsored by Turkish Academy of Sciences under the Young Scientist Award Programme.  




\section*{References}
\bibliographystyle{ifacconf} 
\bibliography{CPHSmodeling}


%
%

\end{document}